\documentclass[aps,prx,twocolumn,superscriptaddress,showpacs,floatfix]{revtex4-2}
\usepackage{comment}
\usepackage{physics}
\usepackage{amsmath,amsthm,amssymb,amsfonts, color, comment, graphicx}
\usepackage{xcolor}
\usepackage{float}
\usepackage{graphicx}
\usepackage{dcolumn}
\usepackage{bm}
\UseRawInputEncoding 
\usepackage{dsfont}
\usepackage{hyperref}

\usepackage[normalem]{ulem}

\newcommand{\bea}{\begin{eqnarray}}
\newcommand{\eea}{\end{eqnarray}}

\newcommand{\bS}{\mathbf{S}}

\newcommand{\br}{\mathbf{r}}

\newcommand{\bod}{\boldsymbol{\delta}}
\newcommand{\bor}{\mathbf{r}}
\newcommand{\bok}{\mathbf{k}}
\newcommand{\boq}{\mathbf{Q}}

\newcommand{\be}{\begin{equation}}
\newcommand{\ee}{\end{equation}}
\newcommand{\bk}{{{\bf{k}}}}

\newcommand{\beal}{\begin{align}}
\newcommand{\eeal}{\end{align}}
\newcommand{\ra}{\rangle}
\newcommand{\la}{\langle}

\newcommand{\upa}{\uparrow}
\newcommand{\dna}{\downarrow}

\newcommand{\dg}{{\dagger}}
\newcommand{\pdg}{{\phantom\dagger}}

\newcommand{\btjstrw}{\mathrel{{\rotatebox[origin=c]{90}
{$\bowtie$}}\kern-0.18em\raisebox{-.95ex}{$\bullet$}
\kern-0.5em\raisebox{.97ex}{$\bullet$}
\kern-1.12em\raisebox{.97ex}{$\bullet$}
\kern-0.52em\raisebox{-.95ex}{$\bullet$}}}

\newcommand{\btjnbrR}{{\mathrel{\rotatebox[origin=c]{90}
{$\bowtie$}}\kern-0.22em\raisebox{.9ex}{$\bullet$}
\kern-1.em\raisebox{-.8ex}{$\bullet$}}}
\newcommand{\btjnbrL}{{\mathrel{\rotatebox[origin=c]{90}
{$\bowtie$}}\kern-0.22em\raisebox{-.8ex}{$\bullet$}
\kern-1.em\raisebox{+.9ex}{$\bullet$}}}

\def\a{\alpha}
\def\b{\beta}
\def\c{\chi}
\def\d{\delta}
\def\e{\epsilon}

\def\g{\gamma}

\def\m{\mu}
\def\n{\nu}
\def\p{\pi}

\def\s{\sigma}
\def\t{\tau}

\def\w{\omega}

\def\D{\Delta}

\def\G{\Gamma}
\def\W{\Omega}

\def\mh{{\mathcal{H}}}
\def\mi{{\mathcal{I}}}
\def\mj{{\mathcal{J}}}

\newcommand{\llangle}[1][]{\savebox{\@brx}{\(\m@th{#1\langle}\)}%
  \mathopen{\copy\@brx\kern-0.5\wd\@brx\usebox{\@brx}}}
\newcommand{\rrangle}[1][]{\savebox{\@brx}{\(\m@th{#1\rangle}\)}%
  \mathclose{\copy\@brx\kern-0.5\wd\@brx\usebox{\@brx}}}

\newcommand{\alm}{{$A\ell M$}}

\usepackage{graphicx}
\usepackage[caption=false]{subfig}

\begin{document}

\preprint{APS/123-QED}

\title{Altermagnetism and superconductivity in a multiorbital $t$-$J$ model}

\author{Anjishnu Bose}
\thanks{These authors contributed equally to this work.}
\author{Samuel Vadnais}
\thanks{These authors contributed equally to this work.}
\author{Arun Paramekanti}
\affiliation{Department of Physics, University of Toronto, 60 St. George Street, Toronto, ON, M5S 1A7 Canada}
\email{arun.paramekanti@utoronto.ca}

\date{\today}

\begin{abstract}
Motivated by exploring correlated multi-orbital altermagnets ({\alm}s) we study minimal $t$-$J$ models on
the square-octagon lattice which 
favors such a collinear magnetic order. While antiferromagnetic order breaks translational and
time-reversal symmetries, the {\alm} state (equivalently, a `$d$-wave ferromagnet') 
features multipolar order which separately breaks time-reversal and 
crystal rotation symmetries but preserves their product leading to spin-split bands with zero net magnetization.
\textcolor{black}{We study the mean field phase diagram of these multiorbital models as we vary doping and interactions, discovering
two types of {\alm} order: (i) itinerant weak-coupling {\alm} metals driven by quasi-1D van Hove singularities, as well as 
(ii) strong {\alm} order at half-filling.
We also find regimes of superconductivity including uniform $s$-wave and $d_{xy}$-wave pairing states,
incipient $d_{xy}$-wave pair density wave order, and uniform
phases with coexisting singlet-triplet pairing and {\alm} order. Our inhomogeneous mean field theory approach
reveals that the coexistence phases are unstable to phase
separation, but longer-range interactions could lead to stripe order.} Our results may be relevant to
studies of doping and pressure on {\alm} materials.
\end{abstract}

\maketitle




\section{Introduction}

The one-band Hubbard 
and $tJ$ models 
have been extensively investigated following the discovery of high temperature cuprate superconductivity
\cite{Hubbard_review_ARCMP2022,cuprate_review_dagotto_rmp1994,cuprate_review_lee_rmp2006,cuprate_rvbreview_2004}.
These models are thought to capture the physics of a wide variety of correlated superconductors which descend 
from nearby states with Mott insulating antiferromagnetic order or other collinear spin density wave orders
\cite{cuprate_review_lee_rmp2006, FeSCAFM, FeSCMagFrust, SDW_Dong_2008, SDW_RevModPhys.60.209}. Indeed, 
fluctuating magnetism is thought to be 
the `pairing glue' responsible for unconventional superconductivity \cite{scalapino_superconductivity_1999,spinflucn_abanov_epl2001,Fe_sc_chubukov2008,Fe_sc_review_chubukov2012}.
In this context it is interesting to explore the effect of doping
on materials with more complex collinear magnetic orders such as the recently discovered altermagnet ({\alm}) order \cite{AMReview_PhysRevX.12.040501, AM_PhysRevX.12.031042}. Altermagnets are 
naturally multi-orbital systems, which host magnetic order that is akin to a type of multipolar order \cite{Multipolar_PhysRevX.14.011019}. {\alm}s preserve translational
symmetry unlike AFMs. While they break time-reversal and lattice rotation symmetries,
they preserve the product of these two symmetries which leads to zero net magnetization while still hosting spin-split bands 
\cite{ALM_Nematic_KivelsonRMP_2003,ALM_Nematic_RaghuPRB_2003,ALMSpinSplit_PhysRevLett.126.127701,SpinSplit_Satoru2019}. This
makes {\alm}s of potential interest for spintronic applications \cite{Spintronics_Jungwirth2016, SpinHall_RevModPhys.87.1213}.
and several candidate materials such as CrSb \cite{ALM_materials_CrSb2024}, Mn$_5$Si$_3$ \cite{ALM_materials_mn5si3_2024anisotropy}, 
MnTe \cite{ALM_materials_MnTe2024}, \(\kappa\)-Cl \cite{kappa_chloride}, and many others have been experimentally explored.
Recent work has explore the Landau theory of altermagnets \cite{mcclarty2023landau}, their symmetry classification 
and their nodal
excitations \cite{schiff2023spin,antonenko2024mirror,fernandes_prb2024}, and coupling of magnetism
to phonons \cite{mcclarty2023landau,steward_dynamic_2023}. Electrons which are proximity coupled to magnetic fluctuations in such altermagnets can potentially form
$p$-wave triplet superconductors \cite{AM_Brekke2023TwodimensionalAS,tripletAM_Brataas2024,tripletAM_Zhang2024,maeland2024manybody}, 
while
proximity coupling of {\alm} to conventional superconducting states or superconductivity induced by local
pairing interactions could support pair density wave states and 
topological superconductivity \cite{ALM_PDW_Fradkin2014,ALM_topologicalSC_Zhu2023,majorana_Flensberg2021}
with edge or corner Majorana modes \cite{majorana_Flensberg2021, ghorashi2023altermagnetic}, and {\alm} order can mediate unconventional Josephson 
effects \cite{ALM_Josephson_Linder2023}, or diode effects \cite{banerjee2024altermagnetic}. In this context, it is interesting to explore
possible superconductivity in doped or pressurized {\alm} ordered materials.

A second important ingredient of superconductivity in many of the correlated quantum materials is that they exhibit
multiple bands, at or near the Fermi level,
which emerge from multiple atomic orbitals or multiple atoms in the unit cell. Such multiorbital models are important for 
a careful microscopic modelling of the CuO$_2$ layers in the cuprate superconductors 
\cite{cuprate_emery1987,cuprate_emery1988,cuprate_varma1997,cuprate_Fischer2014,Cuprate4band,CuprateMai2021,Cuprateselfcons}, 
and play a more direct role in 
superconductors like Sr$_2$RuO$_4$ \cite{Sr2RuO4Multi, Sr2RuO4d+ig, Sr2RuO4SpinSuscep, Sr2RuO4Suzuki2023} and the more recently discovered iron-based \cite{IronSCSi2016} and nickel-based high temperature
superconductors \cite{nickelate_hwang_2019,nickelate_bilayer_wang_2023,nickelate_kuroki_prl2020,nickelate_wu_2019,nickelate_sahadasgupta2020,nickelate_thomale_2020,nickelate_fuchunzhang_prb2020,nickelate_werner_prb2020,nickelate_yahuizhang_2020}.
Multi-orbital models also leads to a more natural description of the rich variety of orders
found generically in these systems including stripe and nematic orders \cite{nematicBaek2015, nematicGlasbrenner2015, StripeWang2016, nematicWang2016}. It is thus interesting to ask if multiorbital systems can also
more naturally host unusual superconductivity associated with nonzero-momentum pairing called `pair density wave order' which have been
found in numerical studies of simple phenomenological models
\cite{PDW_annurev-conmatphys-031119-050711,pdw_paramekanti2010,pdw_jiang2023,PDW_PhysRevLett.130.026001,PDW_Setty2023,pdw_ticea2024}.

Motivated by both the above sets of observations, our work in this paper explores $tJ$ models in a multiorbital system. We consider
a square-octagon lattice model which can support either AFM or {\alm} order depending on the sign of certain exchange interactions; such a lattice has been explored in previous theoretical work \cite{square_octagon_swaveSC, square_octagon_flatSC, Wunderlich_2023} has been proposed to be relevant to certain transitional metal dichalcogenide and nitrogen group monolayers \cite{square_octagon_dichalc, ZHANG2015109}. We study this lattice model as an
illustrative model of how magnetic exchange interactions in an {\alm} could potentially drive superconductivity.
The
simplest {\alm} order which we will consider in this paper may be termed a $d$-wave ferromagnet. Starting with this model, we
study the effect of doping and interactions in such $tJ$ models using mean field theory, exploring both uniform symmetry broken states as well as possible 
spatially modulated orders. \textcolor{black}{Our main result is that the resulting phase diagram contains a rich plethora of phases which include 
metallic or insulating phases with strong AFM/{\alm} order at half-filling, 
metallic phases with itinerant weak-coupling {\alm} order or $d$-wave spin density wave states
induced by quasi-one-dimensional van Hove singularities}, and
$s$-wave and $d$-wave paired superconductors. 
We also find regimes of incipient $d$-wave pair density wave states 
over a range of densities at weak coupling, which could potentially be stabilized with additional nonlocal interactions. In addition,
we find states where {AFM}/{\alm} order coexists with $d$-wave superconductivity leading to mixed singlet-triplet pairing, which however
appear unstable to phase separation into droplets with
Mott insulating {AFM}/{\alm} order and droplets with $d$-wave superconductivity - these may form more organized structures (e.g., stripes)
in the presence of longer range interactions.

\section{Multi-orbital Model}

\begin{figure}[t]
\centering
\includegraphics[width=0.3\textwidth]{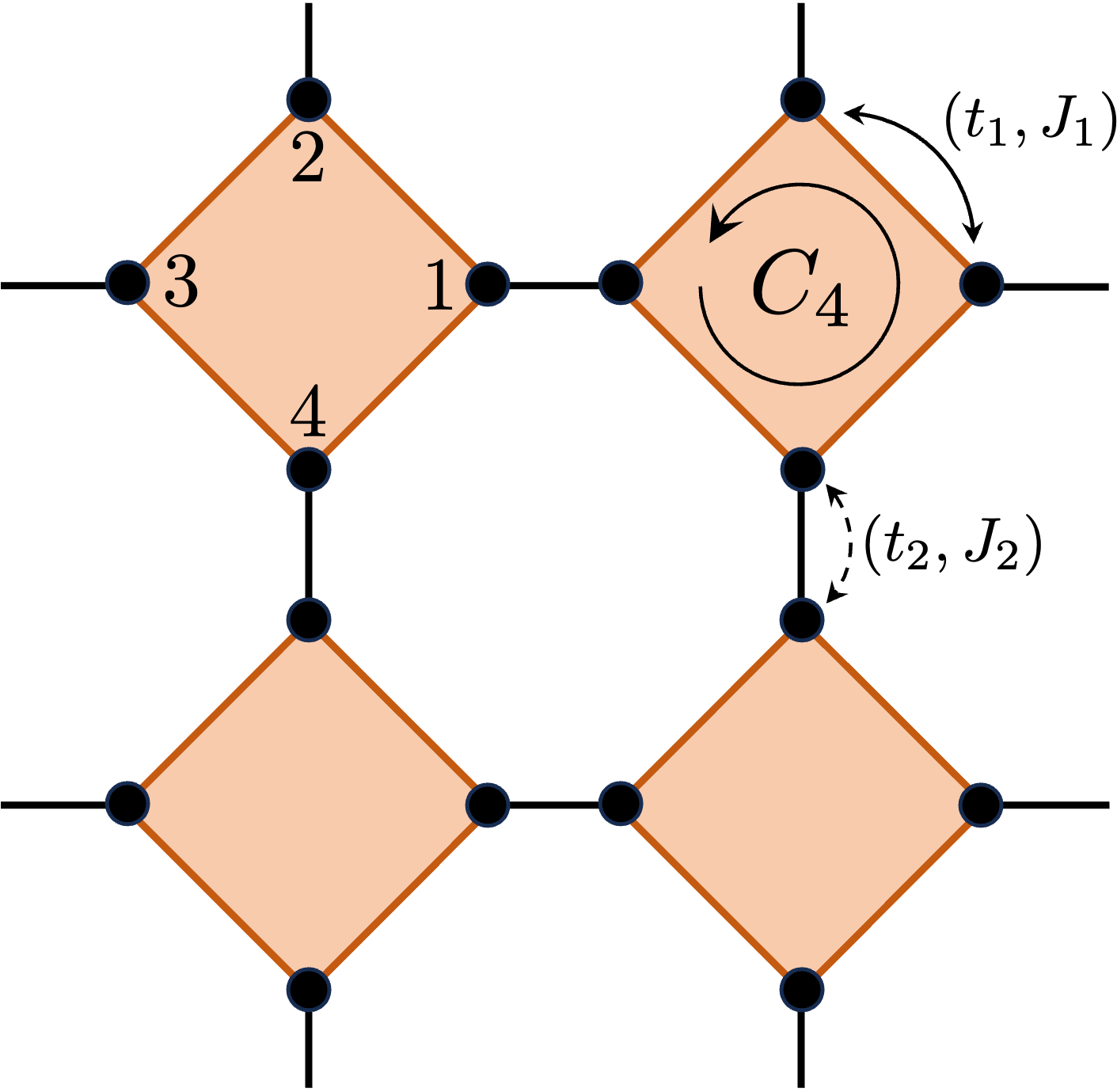}
\caption{Decorated square lattice model, also known as the \emph{square-octagon} lattice, with four-site unit cell. The intra-cell hopping and spin exchange are denoted by $(t_1,J_1)$, and
inter-cell hopping and spin exchange are denoted by $(t_2,J_2)$. We set $t_1\!=\!1$, $t_2\!=\! t_1/2$, and $J_2/J_1\!=\!-J/4$.
We vary the electron filling $\bar{n}$ and interaction $J_1/t$, keeping $J_1 \!>\! 0$.}
\label{fig:lattice}
\end{figure}

We consider a decorated square lattice model with four sites
per unit cell as depicted in Fig.~\ref{fig:lattice}. The Hamiltonian we study is
\begin{eqnarray}
\begin{gathered}
\label{tJHam}
    \!\! H = -\!\!\! \sum_{\la i,j \ra,\sigma}\!\! t_{ij} (c^\dg_{i\sigma} c^\pdg_{j\sigma}\!+\! {\rm h.c.})\!+\!\! \sum_{\la i,j \ra} J_{ij} \bS_i \!\cdot\! \bS_j  \!-\! \mu \! \sum_i n^\pdg_{i\sigma}\,.
\end{gathered}
    \label{eq:hamtj}
\end{eqnarray}
Here, the intra-cell and inter-cell nearest-neighbor hoppings are $t_{ij}=t_1$ and $t_{ij}=t_2$ respectively, and the corresponding
spin exchange couplings on those bonds are denoted by $J_{ij}=J_1$ and $J_{ij}=J_2$ respectively. We set $t_1=1$ to define the
unit of energy, fix parameters $t_2/t_1=1/2$, consider different $J_2/J_1$, and tune the electron filling $0 < \bar{n} < 1$
and interaction strength $J_1/t_1$. $\bar{n}=1$ corresponds to $8$ electrons per unit cell. While the $J_2 > 0$ arises naturally
from a Hubbard model, the case $J_2 < 0$ can only arise from additional Hund's coupling physics with multiple on-site orbitals,
so we view it here only
as a convenient model Hamiltonian to capture {\alm} order in this example. 

We are not implementing strict Gutzwiller 
projection of the electrons in this study - we may view these $tJ$ models as having renormalized hopping and interactions in the spirit of
`renormalized mean field theory' \cite{RMFT1988,cuprate_rvbreview_2004}. In this case, the filling dependent
renormalization of hopping and exchange couplings are given by $t \equiv g_t t_{\rm bare}$ and $J \equiv g_J J_{\rm bare}$, where $g_t \!=\! (1-2 \bar{n})/(1-\bar{n})$ and
$g_J\!=\! 1/(1-\bar{n})^2$. This results in the renormalized $J/t \equiv \lambda(\bar{n}) (J/t)_{\rm bare}$ where $\lambda(\bar{n})\!=\! g_J/g_t$. The
above result leads to 
\begin{equation}
\lambda(\bar{n})\! =\! \frac{1}{(1-2\bar{n})(1-\bar{n})}
\label{eq:rmft}
\end{equation}
Alternatively, we may simply view this $tJ$ model as a toy model to explore 
magnetic and pairing instabilities driven by exchange interactions within unrestricted Hartree-Fock-Bogoliubov theory
in the spirit of earlier work on the one-band model for cuprates \cite{cuprate_tj_sau2015}.

\par

This Hamiltonian has lattice symmetries including translation, a \(\mathcal{C}_4\) rotation around the center of the squares, inversion \(\mathcal{I}\) around the center of the squares and octagons, four mirror symmetries \(\mathcal{M}\), and time-reversal \(\mathcal{T}\). The symmetries of the Hamiltonian also include internal symmetries like \(SU(2)\) spin rotation, and a particle-hole symmetry.\par

\subsection{Orbital basis}

\begin{figure}[t]
    \centering
    \includegraphics[width=0.48\textwidth]{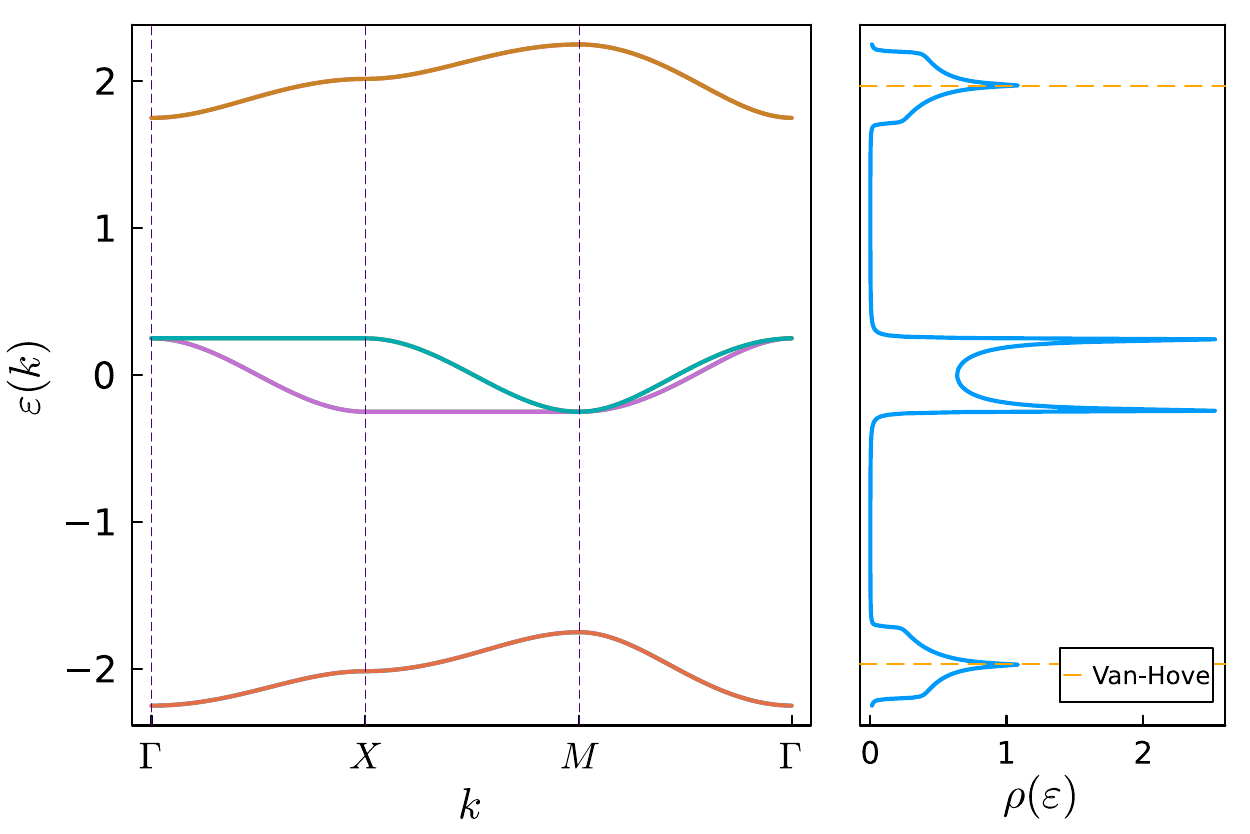}
    \caption{(left) Band structure of the multi-orbital model studied, showing the $s$ (low energy), $X$/$Y$ (intermediate energy), and $d$ (high energy) bands along high symmetry path $\boldsymbol{\Gamma}$-$\boldsymbol{X}$-$\boldsymbol{M}$. The gap between the bands $\sim  2 t_1$, which is evident from the single unit-cell solution. (right) Density of states (DOS) of the non-interacting multi-orbital model. In addition to the $s$-$X/Y$ and $X/Y$-$d$
    band gaps, the DOS shows 2D van Hove singularity (vHS) in the $s$-band ($d$-band)
    bands at filling $n \sim 0.125$ ($n \sim 0.875$) as well as quasi-1D $X/Y$-band vHS at $n \sim 0.26, 0.74$.}
    \label{fig:Band_Structure}
\end{figure}

In the absence of interactions, it is useful to solve the single unit cell in terms of ``orbitals''. We label each site
$i$ in terms of unit cell position $\br$ and basis $\ell=1$-$4$. Using this, the orbitals correspond to
\begin{eqnarray}
    s_{\br \sigma} &=& \frac{1}{2} (c_{\br 1 \sigma} + c_{\br 2 \sigma} + c_{\br 3 \sigma} + c_{\br 4 \sigma})   \label{eq:orbital_s} \\
    d_{\br \sigma} &=& \frac{1}{2} (c_{\br 1 \sigma} - c_{\br 2 \sigma} + c_{\br 3 \sigma} - c_{\br 4 \sigma})   \label{eq:orbital_d} \\
    X_{\br \sigma} &=& \frac{1}{\sqrt{2}} (c_{\br 1 \sigma} - c_{\br 3 \sigma})   \label{eq:orbital_px} \\
    Y_{\br \sigma} &=& \frac{1}{\sqrt{2}} (c_{\br 2 \sigma} - c_{\br 4 \sigma})   \label{eq:orbital_py}
\end{eqnarray}
which are respectively $s$-orbital, $d$-orbital, and a pair of degenerate $p$-orbitals $X,Y$. The corresponding energies
for a single unit cell are $\varepsilon_s=-2 t_1$, $\varepsilon_X=\varepsilon_Y=0$, $\varepsilon_d=+2t_1$. These orbitals will form corresponding bands, which for $t_2 \ll t_1$ are non-overlapping. The resulting band structure is shown  on the left side panel, in Figure\ref{fig:Band_Structure}. In this case, with increasing filling, we go from filling the $s$-orbital band, to the $X,Y$-orbital bands, and eventually the $d$-orbital band, with gapped band insulators appearing in between when some of these bands are filled. This is also reflected in the density of states(DOS) shown on the right hand side of Figure \ref{fig:Band_Structure}, where 2D van-Hove singularities(vHS) are present in the $s$ and $d$ bands, as well as quasi 1D vHS in the $X/Y$ bands.

\subsection{Fermi surfaces}

\textcolor{black}{We begin by exploring the Fermi surfaces of this square-octagon lattice model at various fillings, both 
in a symmetry-unbroken phase and in the presence of symmetry-breaking {\alm} order.
To study the impact of {\alm} order, we incorporate a translationally invariant but
basis-site dependent Weiss magnetic field
in the unit cell which takes on value $+h$ for basis sites $1,3$  and $-h$ for basis sites $2,4$, which
reflects the broken symmetry of the {\alm}, so
\begin{equation}
    H_{A \ell M,{\rm Weiss}} = h \sum_{\br\ell\alpha} (-1)^\ell c^\dg_{\br\ell\alpha} \sigma^z_{\alpha\beta} c^\pdg_{\br\ell\beta}.
\end{equation}
The strength of this Weiss field will depend on the microscopic details such as exchange interactions and
filling which we will later explore within a self-consistent mean field treatment of exchange interactions - 
for now, we simply choose a fixed value for $h$ to illustrate the impact of {\alm} order on the Fermi surfaces.}

\begin{figure}[t]
    \centering
    \includegraphics[width=0.22\textwidth]{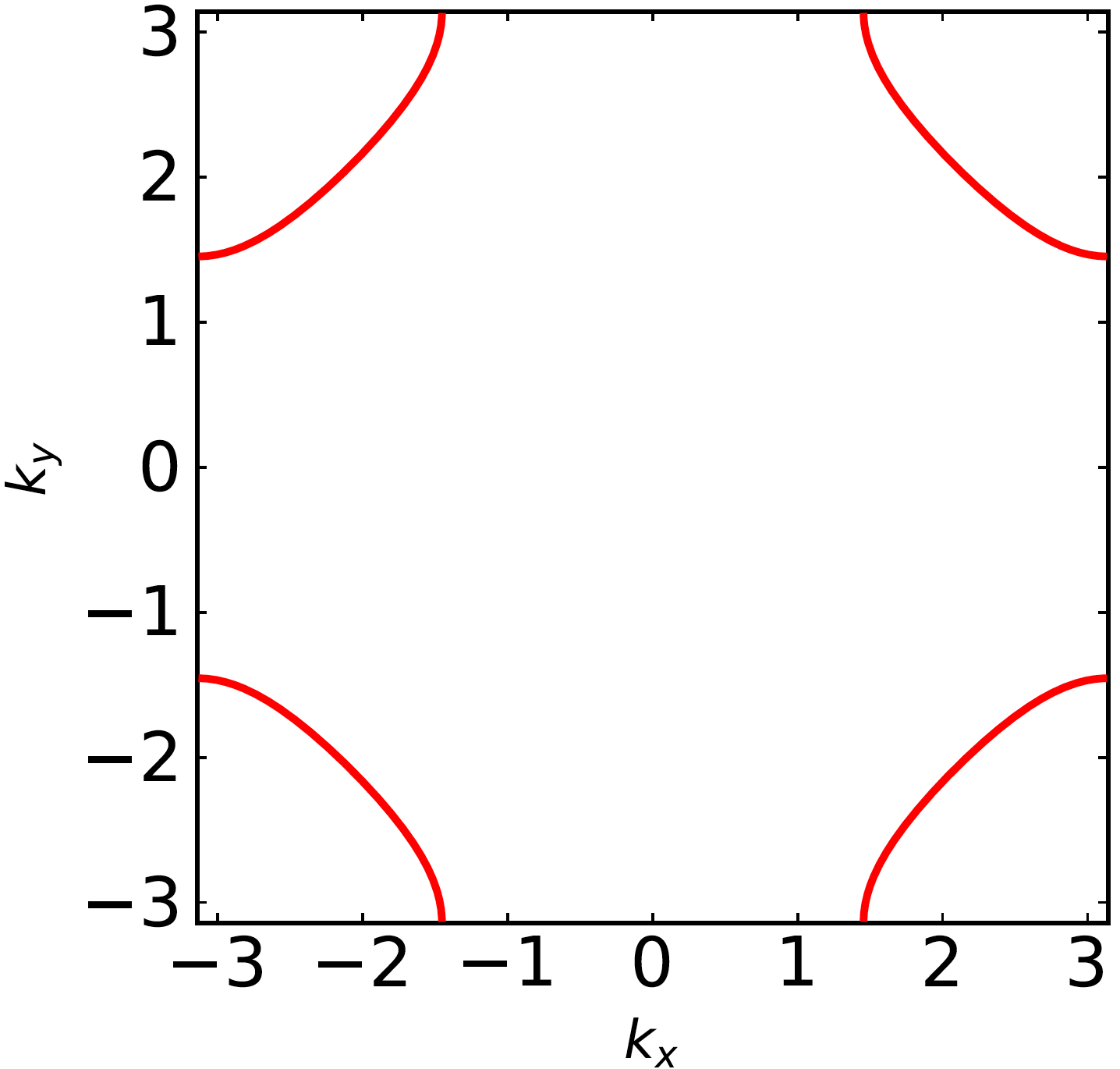}
    \includegraphics[width=0.22\textwidth]{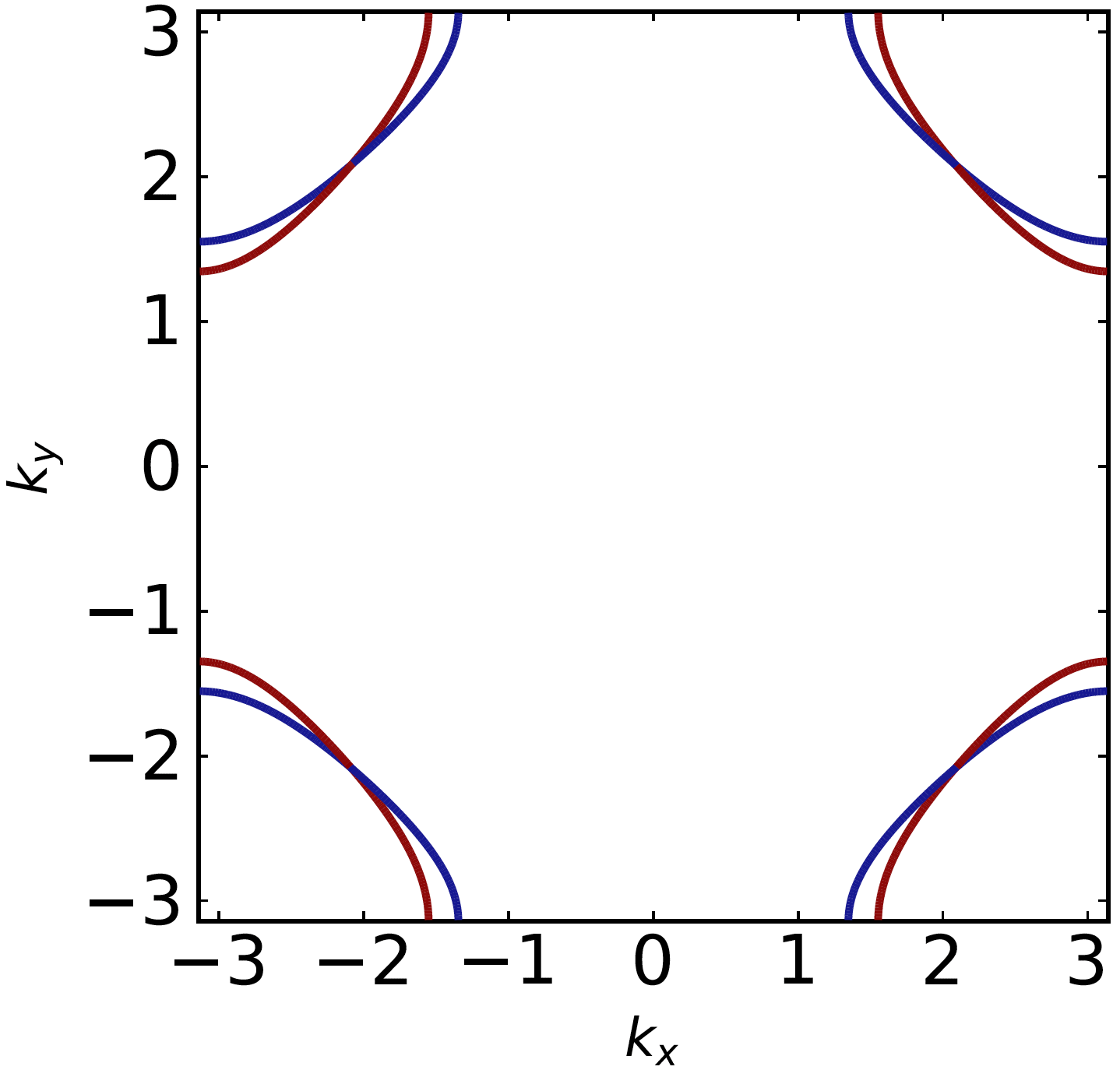}
   \caption{Hole-like Fermi surfaces at $\bar{n}=0.2$ filling where the $s$-orbital band, with dispersion minimum at
    the $\boldsymbol{\Gamma}$ point, is more than half-filled. Left: In the absence of {\alm} order, with spin degenerate FSs.
    Right: \textcolor{black}{Illustrative example of FSs if we incorporate {\alm} order with parameters $t_1 = 1.0$, $t_2 = 0.5t_1$, $h=0.5 t_1$.} The coloring of the bands indicates the spin polarization.}
    \label{FS_n_0_2}
\end{figure}

\begin{figure}[t]
    \centering
    \includegraphics[width=0.22\textwidth]{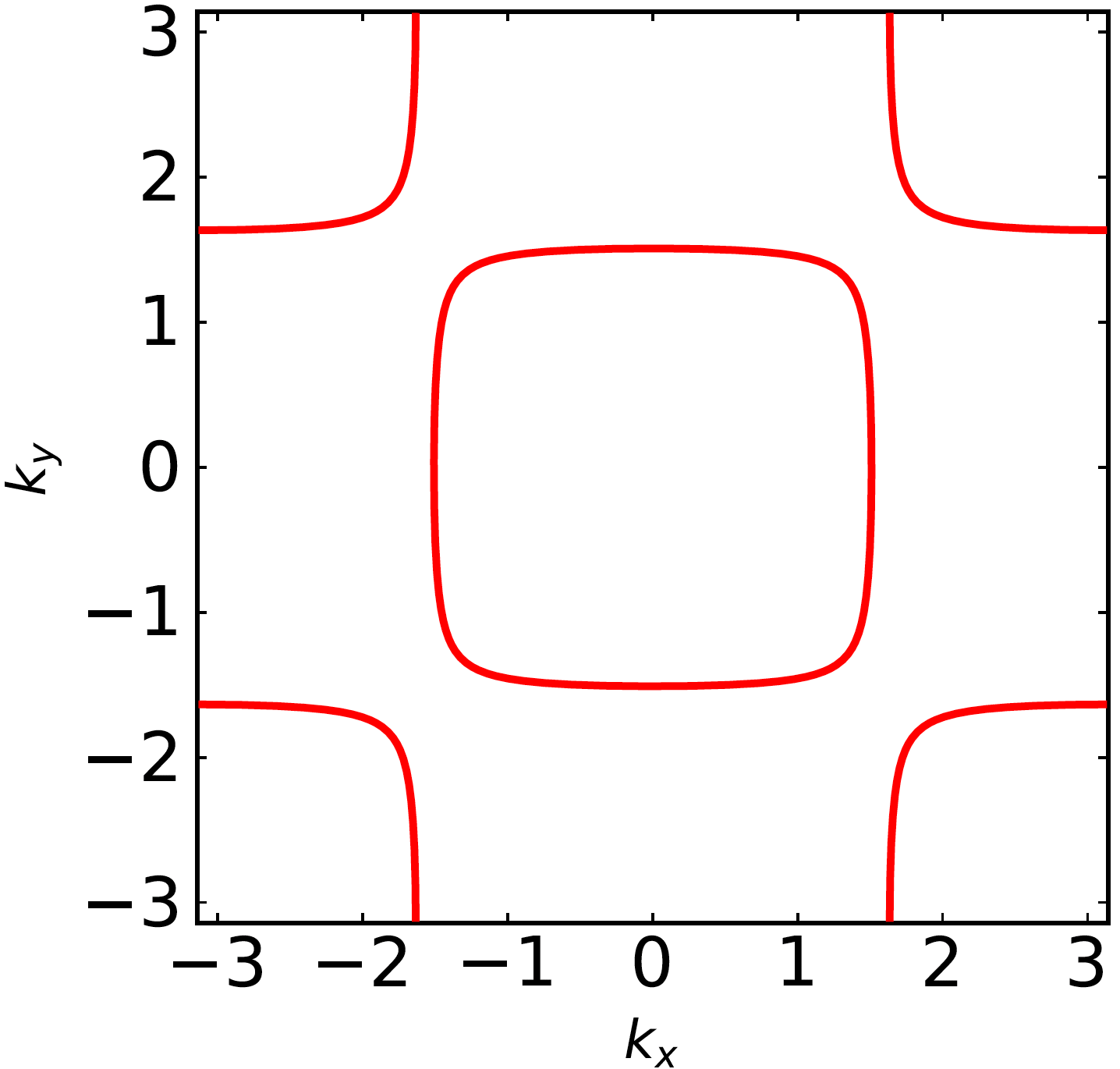}
    \includegraphics[width=0.22\textwidth]{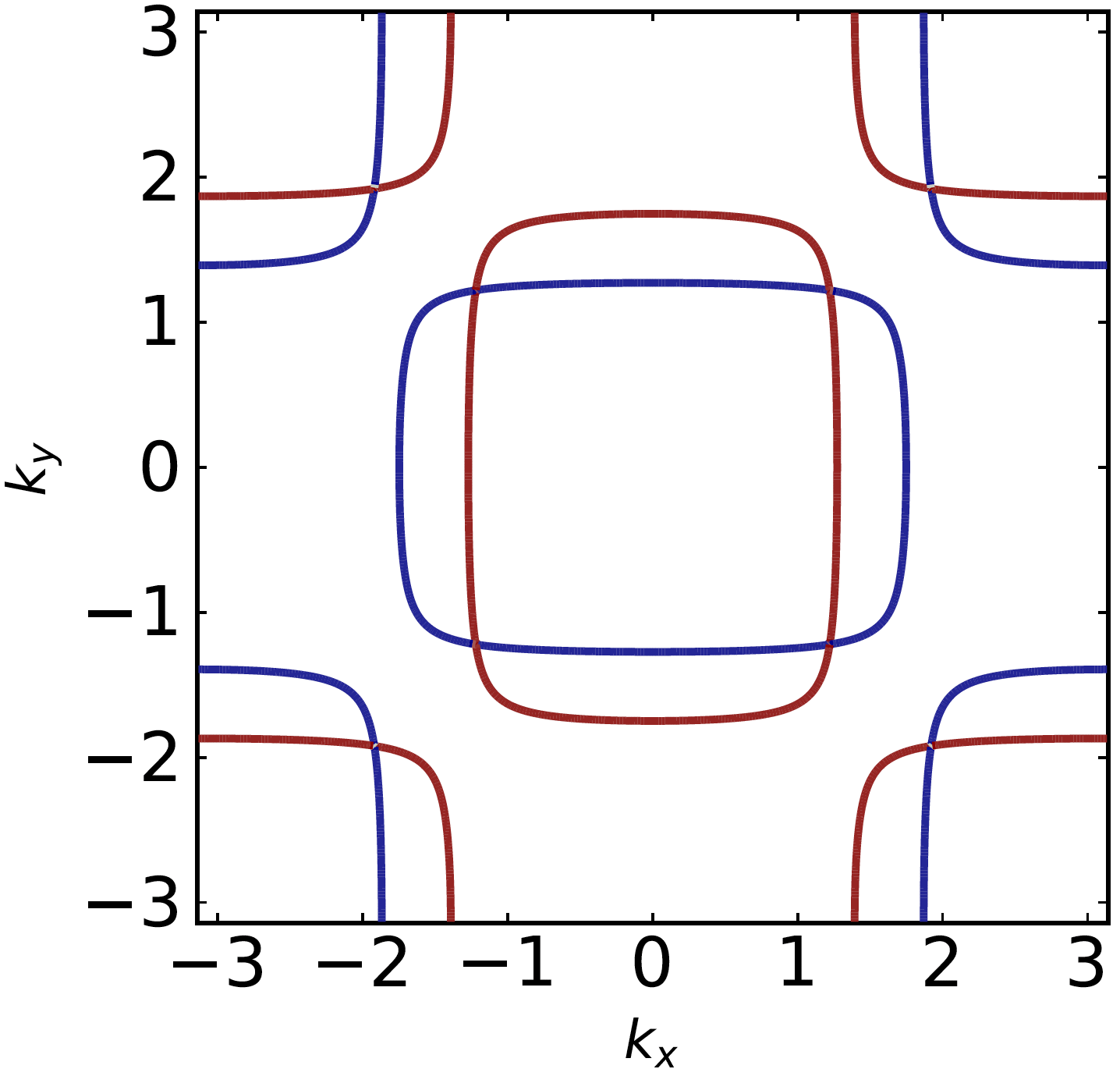}
    \caption{Fermi surfaces at $\Bar{n}$=0.5 filling where the $X,Y$ orbital bands are partially occupied.
    Left: In the absence of {\alm} order, with spin degenerate FSs.
    Right: \textcolor{black}{Illustrative example of FSs if we incorporate {\alm} order with parameters $t_1 = 1.0$, $t_2 = 0.5t_1$, $h=0.5 t_1$.}The coloring of the bands indicates the spin polarization.}
    \label{FS_n_0_5}
\end{figure}

\begin{figure}[t]
    \centering
    \includegraphics[width=0.22\textwidth]{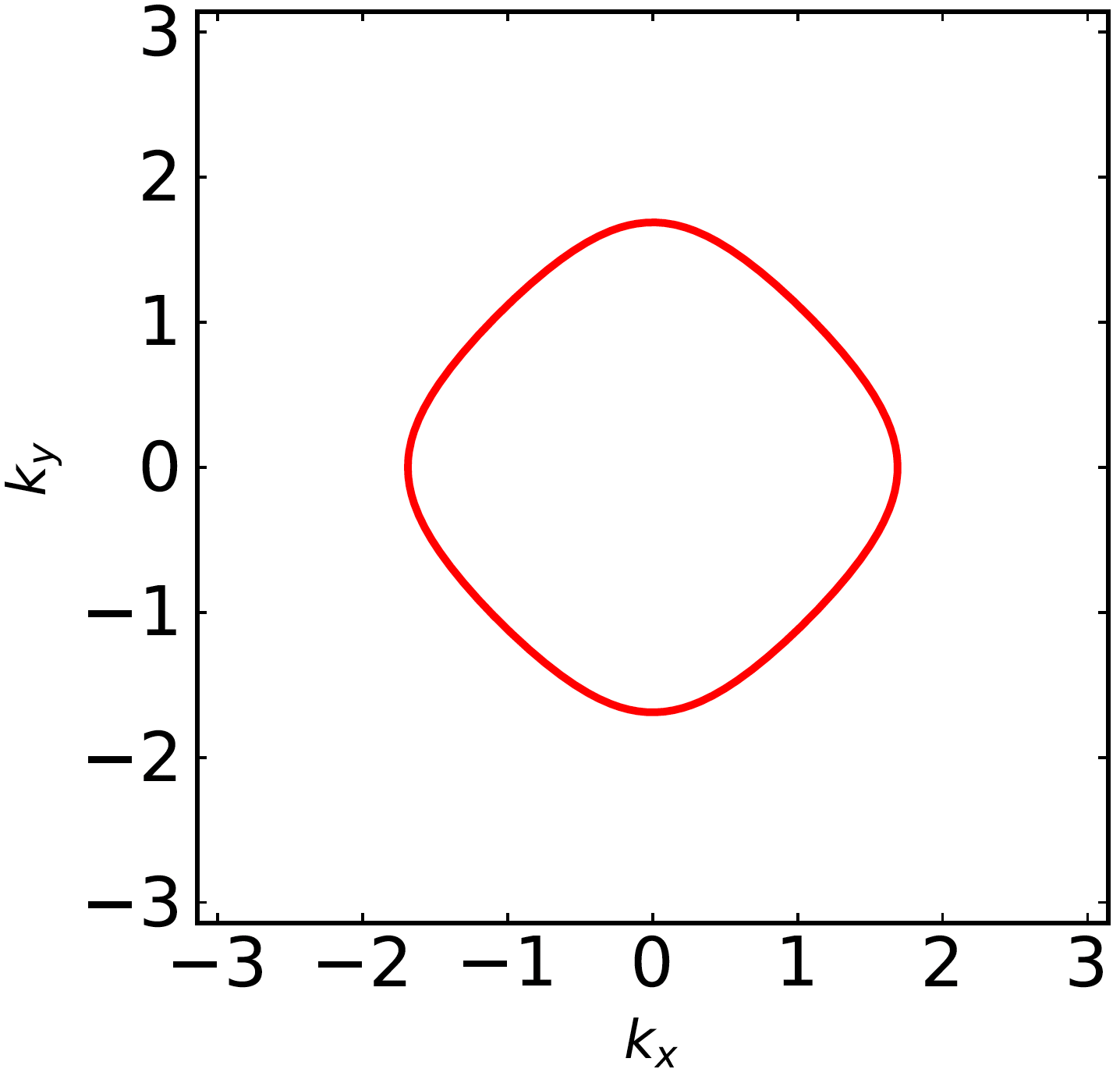}
    \includegraphics[width=0.22\textwidth]{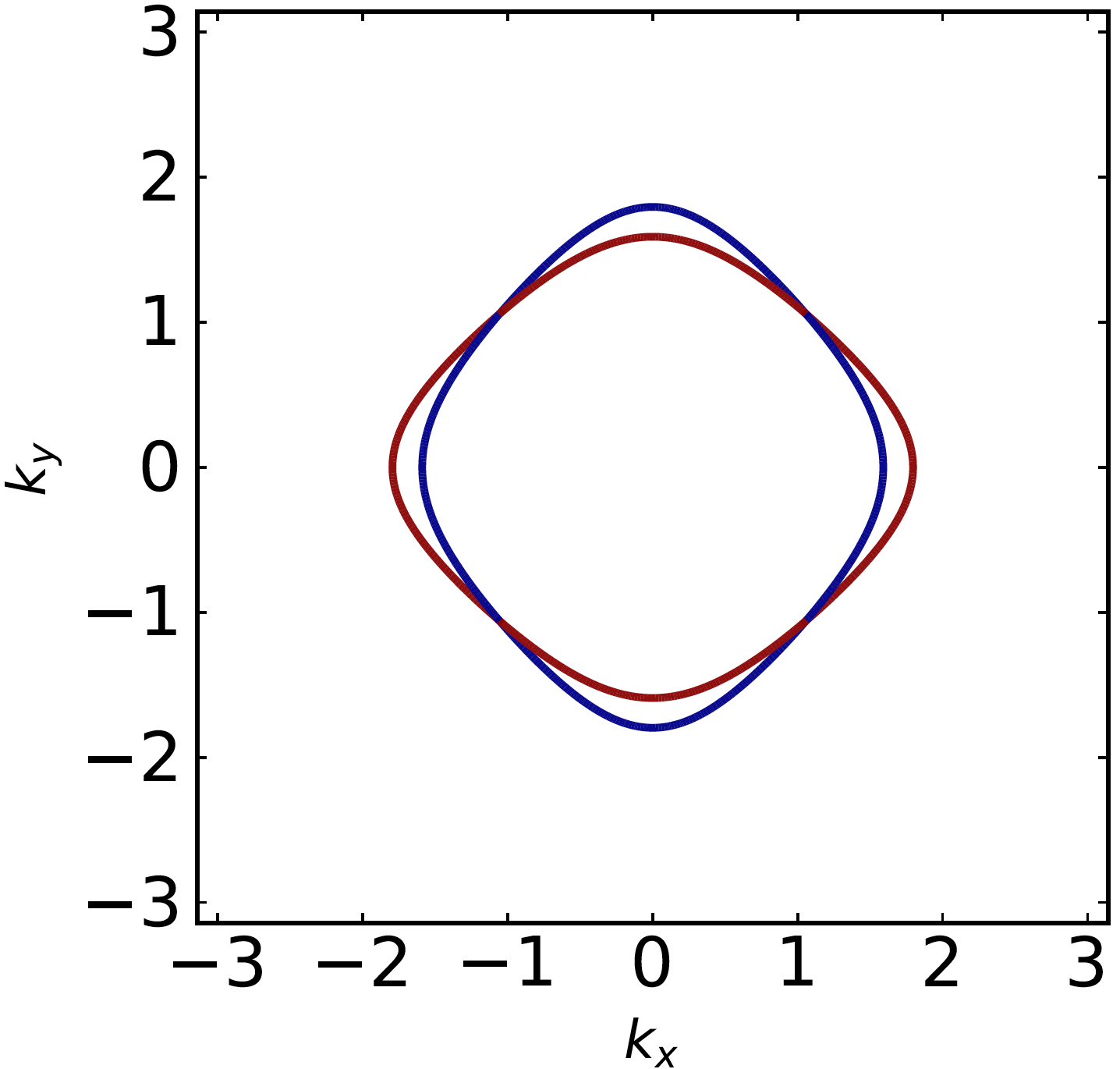}
    \caption{Electron-like Fermi surfaces at $\Bar{n}$=0.8 filling where the $d$-orbital band is partially occupied.
    Left: In the absence of {\alm} order, with spin degenerate FSs.
    Right: \textcolor{black}{Illustrative example of FSs if we incorporate {\alm} order with parameters $t_1 = 1.0$, $t_2 = 0.5 t_1$, $h=0.5 t_1$.} The coloring of the bands indicates the spin polarization.}
    \label{FS_n_0_8}
\end{figure}

\subsubsection{Low filling}
The spin-degenerate Fermi surfaces of the full $8$-band model (including spin) are shown in 
Fig.~\ref{FS_n_0_2} (left) at a filling $\bar{n}=0.2$. To understand this, we note that for 
fillings $\bar{n} < 0.25$, the partially occupied bands arise from the \(s\)-orbitals as given in \eqref{eq:orbital_s}.
This leads to an effective \(s\)-orbitals hopping Hamiltonian which resembles 
a simple square lattice with nearest-neighbour hopping \(t_2\), with an effective 
Hamiltonian
\begin{eqnarray}
H^{\rm eff}_{s} &=& \!\! \sum_{\bk} \! \varepsilon^{s}_\bk  s^\dg_{\bk\sigma} s^\pdg_{\bk\sigma} \\
\varepsilon^{s}_\bk &=& -2 t_1 - \frac{t_2}{2} (\cos k_x + \cos k_y)
\end{eqnarray}
 which has
a minimum at the Brillouin zone (BZ) center $\boldsymbol{\Gamma}=(0,0)$ point. The filling of the $s$-band corresponds to
$n^{\rm eff}_s = 4 \bar{n}$, so that the FS in Fig.~\ref{FS_n_0_2} with $\bar{n}=0.2$ corresponds to 
$n^{\rm eff}_s = 0.8$. 

\textcolor{black}{Incorporating the basis-site dependent {\alm} Weiss field mentioned above leads to
spin-splitting of the FSs as shown in Fig.~\ref{FS_n_0_2} (right).
It is straightforward to show that the basis-site dependent Weiss fields on sites $1$-$4$ 
vanishes if we simply project 
it to the $s$-orbitals on each unit cell, so it has no impact on the FSs. Instead, the observed splitting 
of the $s$-band FS
arises from perturbative mixing of the $s$-band with the Zeeman split $X$ and $Y$ orbital bands, 
which leads to an effective momentum-dependent
Weiss field}
\begin{eqnarray}
H^{A\ell M}_{s} &=& - h \frac{t_2^2}{8 t_1^2}\sum_{\bk,\sigma=\pm} \! (\sin^2 k_x - \sin^2 k_y)  
\sigma s^\dg_{\bk\sigma} s^\pdg_{\bk\sigma}
\end{eqnarray}
\textcolor{black}{As an illustrative example, we set $h=0.5 t_1$.}
This leads to spin splitting at generic momenta on the Fermi surface except along four
`$d$-wave' nodes where the spin-up and spin-down Fermi-surfaces cross when they intersect the \(k_x = \pm k_y\) lines. 
This degeneracy is consistent with the fact that the altermagnet has $x,y$ spatial mirror symmetries and
preserves \(\mathcal{T}\otimes \mathcal{C}_4\).

\subsubsection{Intermediate filling}
Fig.~\ref{FS_n_0_5} shows the FSs at intermediate filling, \(\bar{n} = 1/2\) with and without an {\alm} Weiss field respectively. 
For any filling $0.25 < \bar{n} < 0.75$, 
the bands can be effectively described in terms of an effective hopping Hamiltonian arising from $X$ and $Y$ orbitals hopping along their easy axis (\(x\) and \(y\)) respectively. This leads to a 
\begin{eqnarray}
H^{\rm eff}_{p} &=& \!\! \sum_{\bk} \! \left[\varepsilon^X_\bk  X^\dg_{\bk\sigma} X^\pdg_{\bk\sigma}  \!+\! \varepsilon^Y_\bk
    Y^\dg_{\bk\sigma} Y^\pdg_{\bk\sigma} \right. \nonumber\\
    \!&+&\! \left. \varepsilon^{XY}_\bk (X^\dg_{\bk\sigma} Y^\pdg_{\bk\sigma} \!+\! {\rm h.c.}) \right]
\end{eqnarray}
where $\varepsilon^X_\bk = t_2 \cos k_x$, $\varepsilon^Y_\bk = t_2 \cos k_y$, and an orbital mixing term
with $\varepsilon^{XY}_\bk = t' \sin k_x \sin k_y$. Direct projection does not lead to
this orbital mixing term, but it is generated by perturbative processes which virtually excite to the $s,d$ orbitals
so that $t' \approx t_2^2/2t_1$.
Incorporating the {\alm} Weiss field leads to the additional term
\begin{eqnarray}
H^{A \ell M}_{p} = - h \sum_{\bk\sigma=\pm} \sigma (X^\dg_{\bk\sigma} X^\pdg_{\bk\sigma} - Y^\dg_{\bk\sigma} Y^\pdg_{\bk\sigma}) 
\end{eqnarray}
\textcolor{black}{As an illustrative example, we again set $h=0.5 t_1$.}
The resulting FSs are shown in Fig.~\ref{FS_n_0_5} without and with {\alm} order. In the absence of {\alm} order, we
find two FS pockets, one centered at the $\boldsymbol{\Gamma}=(0,0)$ point and the other at $(\pi,\pi)$.
With {\alm} order, both FS pockets get spin split with $8$
band touching points along the $k_x = \pm k_y$ lines. We note that the {\alm} spin splitting is much larger in this case 
with degenerate $X/Y$-bands at the Fermi level, being $\sim h$, as opposed to the single band case discussed earlier 
where it is $(h t^2_2/t^2_1) \ll h$ due to indirect effects arising from perturbative coupling to the $X/Y$ bands. This might
suggest that multiorbital materials with multiple symmetry related bands at the Fermi level in the paramagnetic metal
are likely to host stronger spin splitting once the system forms an {\alm}.

\subsubsection{High filling}
The simple model we have considered is particle-hole symmetric, invariant under 
$c^\pdg_{\br,i,\sigma} \to (-1)^{\br_x+\br_y} c^\dagger_{\br,i,\sigma}$. This leads to a momentum dispersion 
for the high energy $d$-orbital derived bands at large filling $\bar{n}$,
to be the same as the $s$-orbital band dispersion at low filling $1-\bar{n}$ except for a shift in momentum by 
$(\pi,\pi)$. Fig.~\ref{FS_n_0_8} shows these FSs for $\bar{n}=0.8$ without and with {\alm} order.

\subsection{Projected Interacting Hamiltonians}
We can simplify the full interacting Hamiltonians by projecting to specific orbitals at various fillings. 
For low filling, $0 < \bar{n} < 1/4$, we can project to the $s$-orbital. At intermediate filling, 
$1/4 < \bar{n} < 3/4$, the physics 
is governed by the $X,Y$-orbitals. Finally, at high fillings, $3/4 < \bar{n} < 1$, we project to $d$-orbital 
states. Since the Hamiltonian has particle-hole symmetry, the physics in the $s$-orbital 
dominated regime at low filling $\bar n < 1/4$ is the same as the $d$-orbital regime at high filling
$1-\bar{n} < 1/4$. These simplified models will be useful for interpreting our full numerical
mean field results discussed in the following section. We note that if we use renormalized mean field theory
for the constrained $tJ$ (crudely accounting for projecting out double occupancy via Gutzwiller factors),
then the effective $(J/t)_{\rm eff} = (J/t)_{\rm bare}$

\subsubsection{Low/High filling}
For $\bar{n} < 1/4$, we invert the above equations Eq.~\ref{eq:orbital_s} and then project to the $s$-orbital, 
which we can account for by setting $c_{\br \ell \sigma} \to  s_{\br}/2$ for all sublattices $\ell=1$-$4$. 
Using this, we find
\begin{eqnarray}
    H &=& - \frac{t_2}{4} \!\! \sum_{\la \br,\br' \ra,\sigma} ( s^\dg_{\br\sigma} s^\pdg_{\br'\sigma}
    + {\rm h.c.}) \!-\! \mu \! \sum_{\br\sigma} s^\dg_{\br\sigma} s^\pdg_{\br\sigma} \nonumber \\ 
    &-& \frac{3 J_1}{8} \sum_\br n^{(s)}_{\br\upa} n^{(s)}_{\br\dna} + \frac{J_2}{16} \sum_{\la \br,\br' \ra}
    \bS_{\br} \!\cdot\! \bS_{\br'}
    \label{eq:projS}
\end{eqnarray}
where $n^{(s)}_{\br\sigma} = s^\dg_{\br\sigma} s^\pdg_{\br\sigma}$ and $\bS_\br$ is the $s$-fermion
spin operator.
The intra-cell exchange interactions and inter-cell hopping 
thus lead to an effective square lattice {\em attractive} Hubbard model with 
$U_{\rm eff}/t_{\rm eff} = 3 J_1/2 t_2$
which will favor $s$-wave superconductivity as observed in our mean field calculations. 
Weak ferromagnetic inter-site exchange interaction $J_2$ is not expected to qualitatively
modify the physics of this superconductor.

For high filling, $\bar{n} > 3/4$, the projected model looks identical due to a particle-hole symmetry. The
only difference is that we should replace 
$s_{\br\sigma} \to d_{\br\sigma}$ in Eq.~\ref{eq:projS} for the fermion operators.

\subsubsection{Intermediate filling}
For $1/4 < \bar{n} < 3/4$, we can similarly project the full Hamiltonian to the $X,Y$-orbitals, 
setting $(c_1, c_2, c_3, c_4) \to (X, Y, −X, −Y )/\sqrt{2}$, where we have sup-
pressed site and spin labels for convenience.
This leads to the effective Hamiltonian
\begin{eqnarray}
    H &=& \!\! \sum_{\bk} \! \left[\varepsilon^X_\bk  X^\dg_{\bk\sigma} X^\pdg_{\bk\sigma}  \!+\! \varepsilon^Y_\bk
    Y^\dg_{\bk\sigma} Y^\pdg_{\bk\sigma} \!+\! \varepsilon^{XY}_\bk (X^\dg_{\bk\sigma} Y^\pdg_{\bk\sigma} \!+\! {\rm h.c.}) \right]
    \nonumber \\
&+& J_1\sum_\br \bS^X_{\br} \!\cdot\! \bS^Y_{\br}
    \!-\! \mu \! \sum_\br (n^{X}_{\br\sigma}+n^{Y}_{\br\sigma}) \nonumber \\
&+& \frac{J_2}{4} \sum_{\br} (\bS^X_{\br} \!\cdot\! \bS^X_{\br+\hat{x}}+\bS^Y_{\br} \!\cdot\! \bS^Y_{\br+\hat{y}}) 
\label{eq:projXY}
\end{eqnarray}
where $\varepsilon^X_\bk = t_2 \cos k_x$ and $\varepsilon^Y_\bk = t_2 \cos k_y$.
The operators $\bS^X_{\br},\bS^Y_{\br}$ denote the spin operators
for $X,Y$ orbitals respectively; similarly $n^X_\br,n^Y_\br$ denote number of electrons in $X,Y$ orbitals respectively. The
effective model thus is a two-orbital model with {\em effective AFM Hund's coupling} set by the exchange $J_1$, and a 
much weaker intersite same-orbital exchange coupling $J_2/4$.

\section{Mean field phase diagram}

\begin{figure*}[t]
\centering
\includegraphics[height=0.31\textwidth]{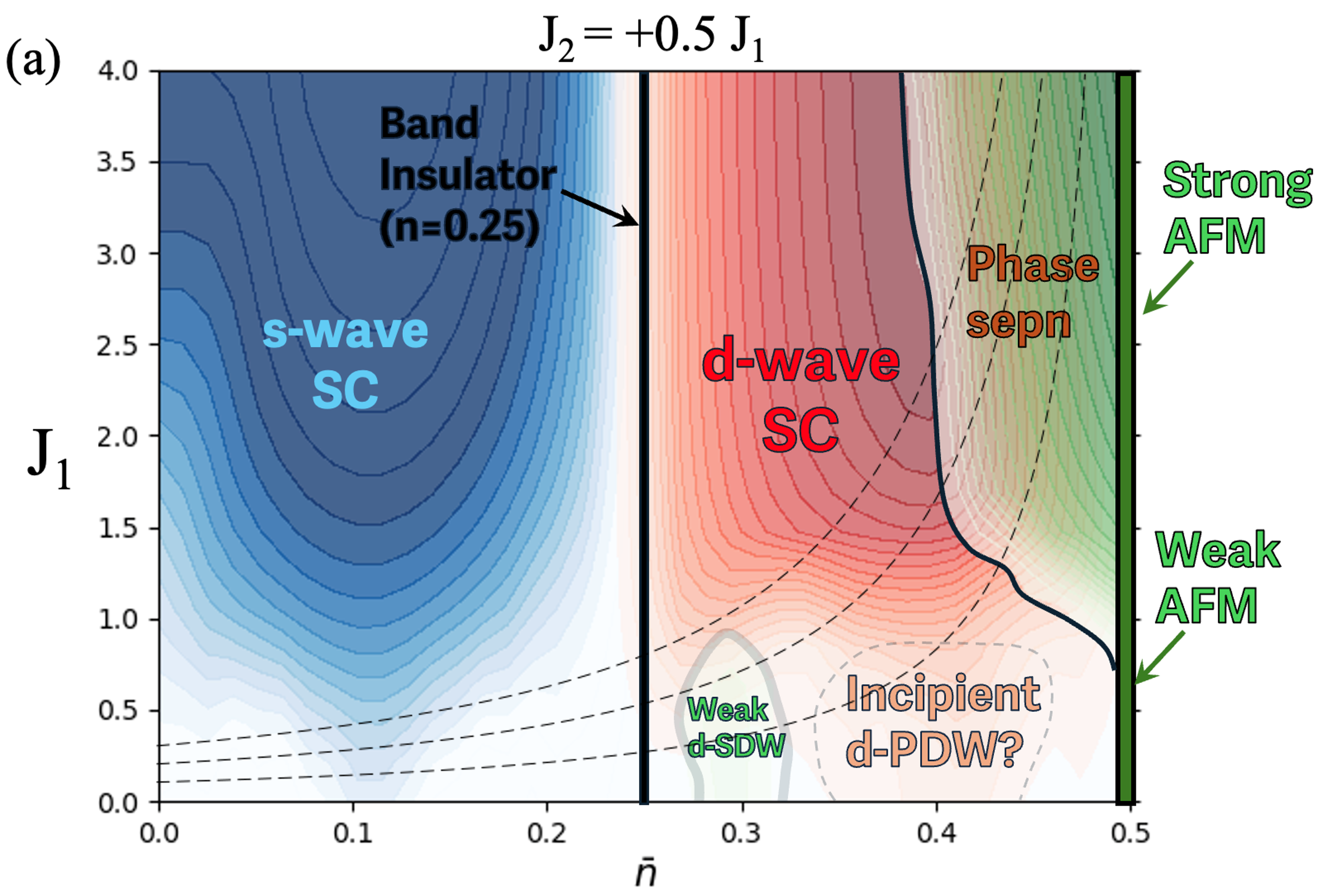}
\includegraphics[height=0.31\textwidth]{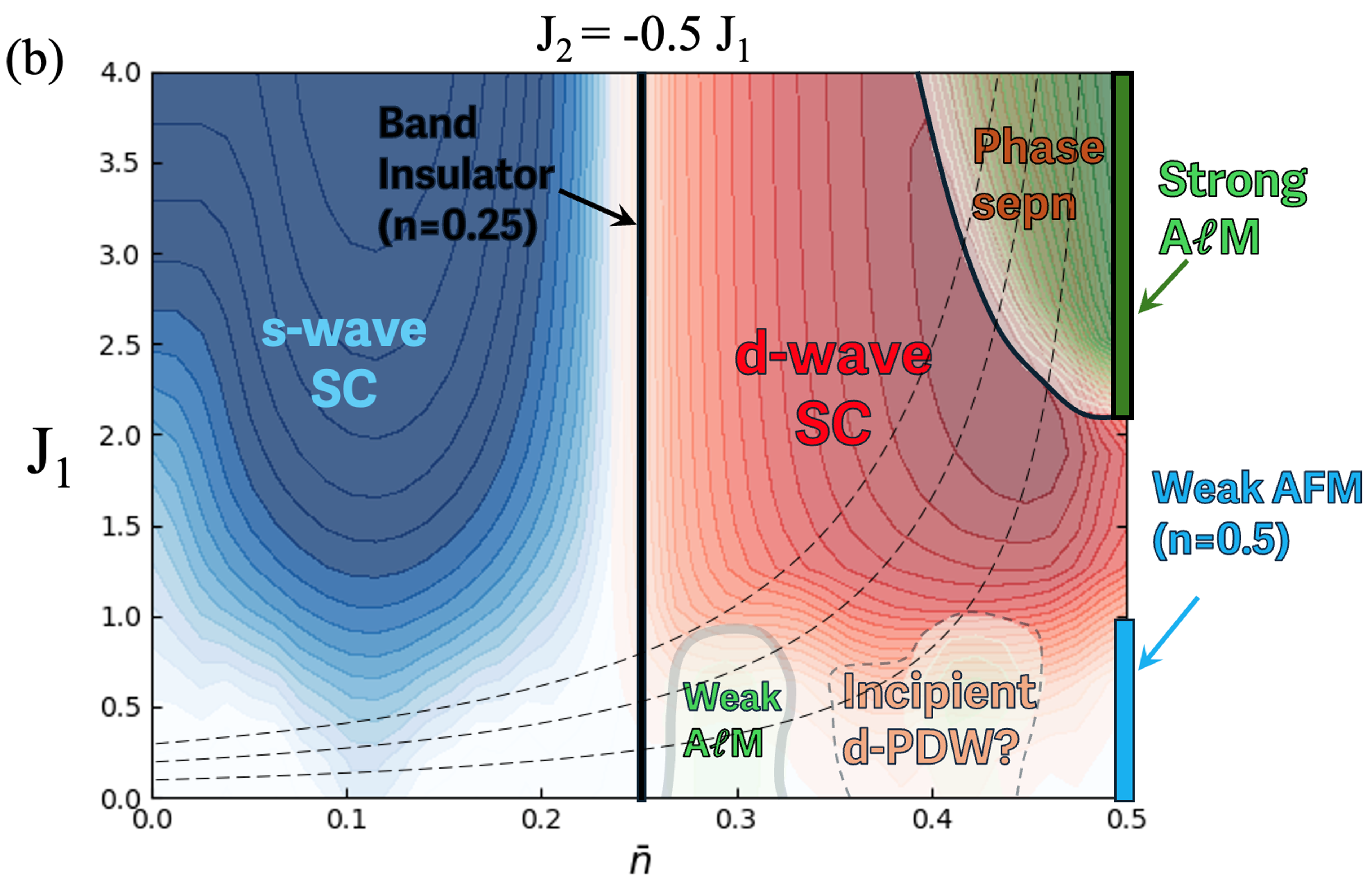}
\caption{Mean field phase diagram of Hamiltonian in Eq.~\ref{eq:hamtj} with
fixed $t_1=1$, $t_2 = 0.5 t_1$, and inter-cell exchange coupling
(a) $J_2 = +0.5 J_1$ and (b) $J_2 = -0.5 J_1$. \textcolor{black}{Dashed lines indicate the renormalized exchange $\lambda(n) J_1$, 
with $\lambda(n)$ defined in Eq.~\ref{eq:rmft}, for
$J_1=0.1,0.2,0.3$ which is applicable if this model is viewed as a renormalized mean field $tJ$ model of Gutzwiller projected electrons.}
In both models,  $J_2 > 0$ and $J_2 < 0$ we find regimes of $s$-wave and $d$-wave superconductors with color contours 
indicating qualitative variation of the pair order parameter. Over a range of densities, the quasi-1D van Hove singularity
leads to weak uniform {\alm} order for $J_2 < 0$, and a modulated
$d$-spin density wave ($d$-SDW) order with {\alm}-type order within 
the unit cell for $J_2 > 0$.
In addition, we find a window at
weak coupling labelled ``incipient d-PDW'' where small system size calculations seem to favor $(\pi,\pi)$ PDW with $d$-wave order over uniform
$d$-wave SC, but the PDW order decreases with increasing system size (so the uniform $d$-SC wins at large system sizes).
At half-filling and strong coupling, we find insulators with strong magnetic orders of 
AFM ($J_2>0$) or {\alm} ($J_2 < 0$) type.
The region marked `Phase sepn' shows phase separation between
$d$-wave SC and the insulating AFM  or {\alm} order. The 
magnetization and pairing patterns for these phases are
shown in Fig.\ref{fig:pattern}, while the nature of phase separation is shown in Fig.\ref{fig:phasesep}.}
\label{fig:phasediag}
\end{figure*}

We solve the $t$-$J$ model Hamiltonian system in mean field theory. We do this in real space by allowing for all possible mean field channels: 
inhomogeneous hopping (including spin dependent hopping), pairing (both singlet and triplet), and on-site densities and magnetizations. 
The mean field Hamiltonian, on any $(i,j)$ bond, decomposed into these three channels looks like the following 
(all spin indices are being summed over):
\begin{eqnarray}
\!\!\!\!\!    H^{\rm MF}_{ij} \!&=&\! \left(H_{\rm hop} \!+\! H_{\rm loc} \!+\! H_{\rm pair}\right)\,,\\
\!\!\!\!\!    H_{\rm hop} \!&=&\! 
c_{i\a}^\dagger \left(\! -t_{ij}\d^{\a\b}+\frac{J_{ij}}{4} \s_{a}^{\a\g}\s_{a}^{\d\b}\chi_{ji}^{\d \g} \! \right) c_{j\b} \!+\! {\rm h.c.} \\
\!\!\!\!\!    H_{\rm loc} \!&=&\! c_{i\a}^\dagger \left(\! -\mu\d^{\a\b} \!+\! \frac{J_{ij}}{4} \s_{a}^{\a\b}\s_{a}^{\g\d}M_j^{\g \d} \! \right) 
c_{i\b} \!+\! {i \leftrightarrow j} \\
\!\!\!\!\!    H_{\rm pair} \!&=&\! c_{i\a}^\dagger \left(\frac{J_{ij}}{4} \s_{a}^{\a\g}\s_{a}^{\b\d}\Delta_{ji}^{\d \g} \right) 
c^{\dagger}_{j\b} \!+\! {\rm h.c.}
\end{eqnarray}
where $\D_{ij}^{\a \b} \!=\! \expval{c_{i\a}c_{j\b}}$, $\chi_{ij}^{\a \b} \!=\!  \expval{c_{i\a}^\dagger c_{j \b}}$,
$M_{i}^{\a \b} \!=\! \expval{c_{i\a}^\dagger c_{i\b}}$.
We solve for all channels self-consistently in real space, at each site and bond,
allowing for a convergence tolerance of $\sim 10^{-3}$ for each parameter at a temperature $T/t_1=0.001$. 
We start with several different 
random initial conditions to maximize sampling of the full phase space and ensure that the target convergence is reached. The converged states 
are then compared to the uniform solution to confirm their stability. Fig.~\ref{fig:phasediag}(a) and Fig.~\ref{fig:phasediag}(b) show the 
phase diagrams obtained from the mean field solution as we vary density $\bar{n}$ and $J_1$ for fixed $J_2/J_1 = 0.5$ and 
$J_2/J_1 = -0.5$ respectively. We have also indicated dashed lines through the phase diagram which corresponds to the variation of
$J_1$ with density within a renormalized mean field theory framework where the bare $J_{1} \to \lambda(\bar{n}) J_{1}$, with $\lambda(\bar{n})$ 
given by Eq.~\ref{eq:rmft}.
Below we discuss some important aspects of these
phase diagrams. We have also computed the mean field phase diagram in momentum space 
allowing for $1 \times 1$ unit cell ($4$ sites) and $2\times 2$ unit cell ($16$ sites); this phase diagram is shown in 
Appendix \ref{appendix:kSpace}; we find that it is largely in agreement with Fig.~\ref{fig:phasediag} except in regimes which show
phase separation or incipient PDW order.

\begin{figure*}[t]
\centering
\includegraphics[width=0.95\textwidth]{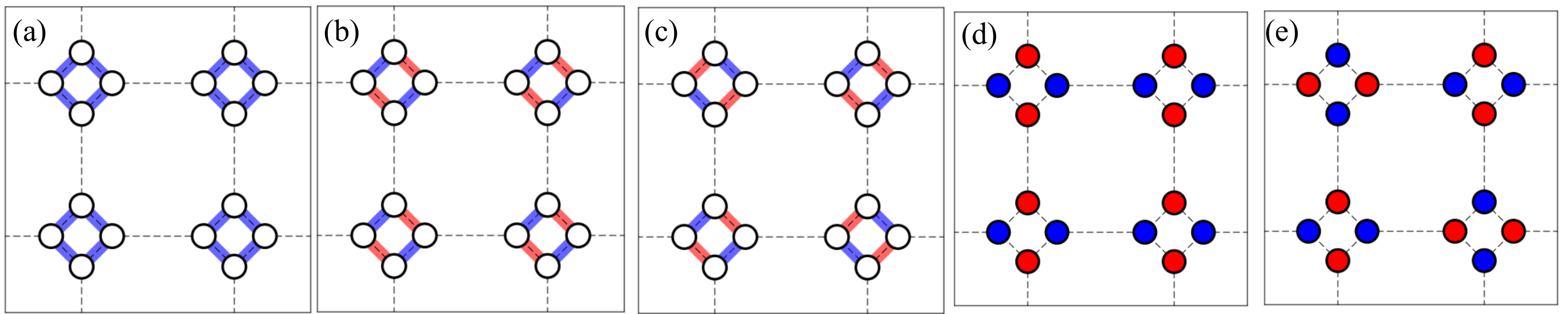}
\caption{Order parameter patterns found in a quadrupled unit cell mean-field theory calculations. (a) \textbf{uniform s-wave pairing} with the same phase and strength of pairing on each intra-diamond bond. (b) \textbf{uniform d-wave pairing} with uniform strength but alternating phase of the pairing within the diamonds. (c) \textbf{d-wave PDW} with a d-wave pattern of pairing within the diamond, but now with a wave-vector of \(\boldsymbol{M} = (\pi, \pi)\). (d) \textbf{Altermagnet} with red marking spin-up and black marking spin-down on site. The altermagnet ordering wave-vector is however simply \(\boldsymbol{\G}\). (e) \textbf{AFM} with similar ordering within the diamond as the altermagnet, but with a wave-vector of \(\boldsymbol{M}\). The weak $d$-SDW is similar to AFM but has modulations at incommensurate
wavevector driven by near nesting of Fermi surface segments.}
\label{fig:pattern}
\end{figure*}

\begin{figure*}[t]
    \centering
    \includegraphics[height=0.25\textwidth]{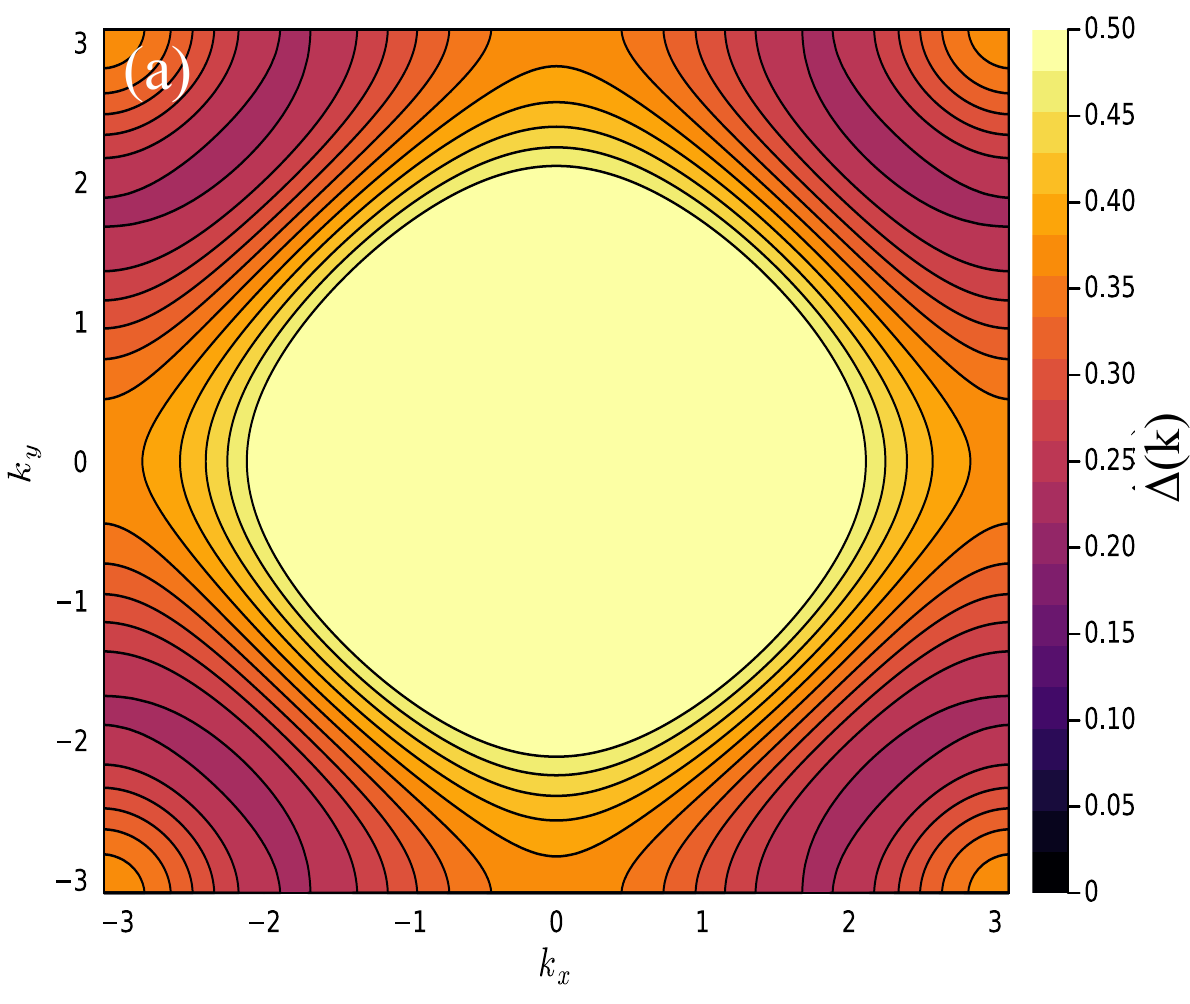}
    \includegraphics[height=0.25\textwidth]{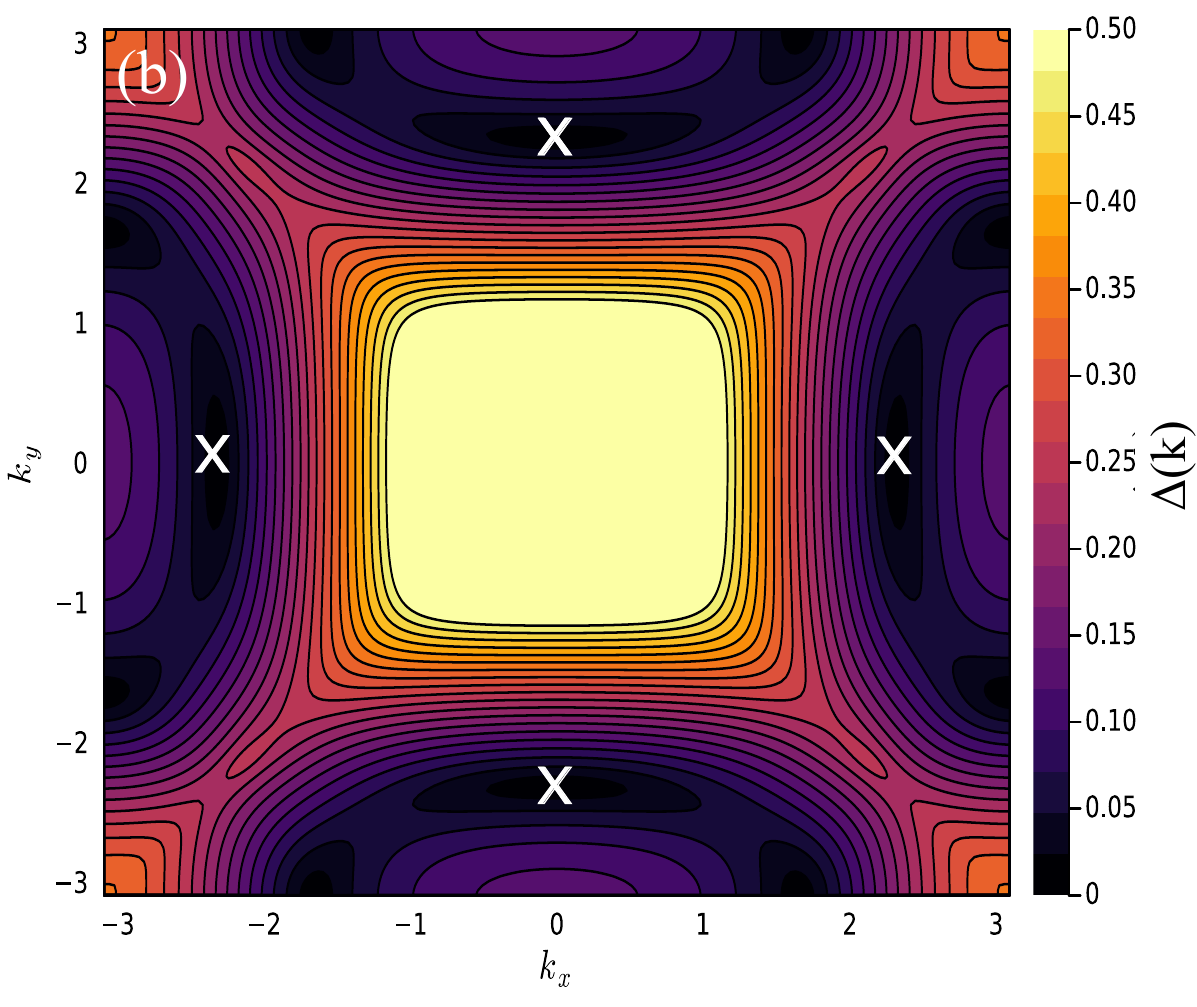}
    \includegraphics[height=0.255\textwidth]{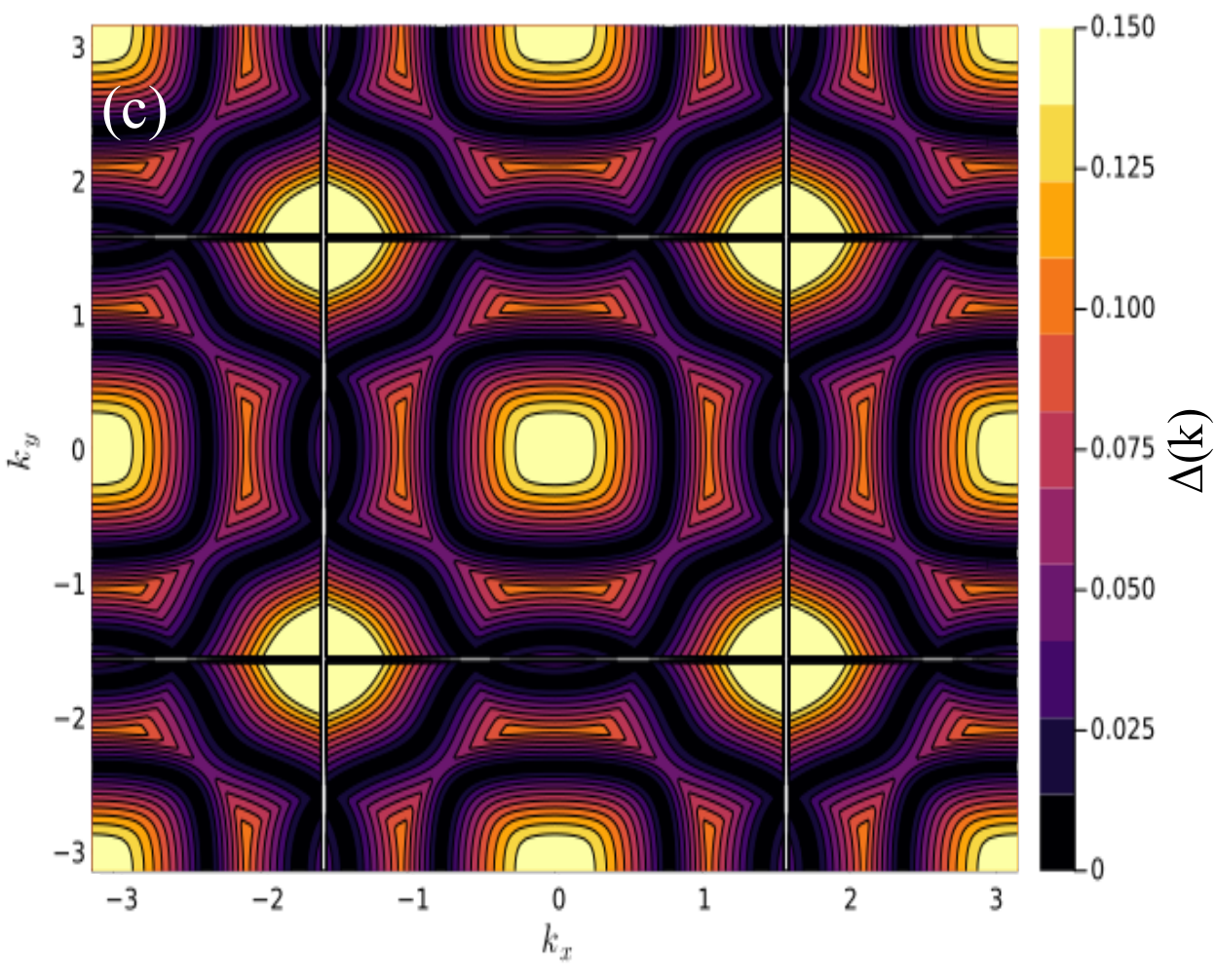}
    \caption{Illustrative plot of the gap in the Bogoliubov quasi-particle spectrum at (a) \(\bar{n} = 0.2\) 
    with s-wave \textcolor{black}{pairing strength} $0.1 t_1$; this refers to the SC pairing Weiss field in the BdG Hamiltonian. 
    This value is realized around $J_1\approx 1.5 t_1$ in the self-consistent BdG
    mean-field theory, and it leads to a fully gapped spectrum.
    (b) \(\bar{n}=0.4\) also with d-wave pairing \textcolor{black}{pairing strength}\(~0.1 t_1\) which shows Dirac nodes (marked with crosses `x') along the \(\boldsymbol{\G}\rightarrow \boldsymbol{X}\), and \(\boldsymbol{\G}\rightarrow \boldsymbol{Y}\) paths in the Brillouin zone. 
    (c) The corresponding case when a $(\pi,\pi)$ d-wave pairing with \textcolor{black}{pairing strength} \(~0.1 t_1\) (this is exaggerated as compared to the mean-field results for illustrative purposes) is considered instead shows a gapless surface of quasiparticle excitations; for comparison
    with (b) we have plotted the spectrum in the original Brillouin zone.}
    \label{fig:spectra}
\end{figure*}

(i) {\bf $s$-wave SC:} In both cases, the low density $\bar{n} < 0.25$
regime shows $s$-wave superconductivity. This is driven by the effective
attractive Hubbard attraction induced by $J_1$ for the $s$-orbital as 
seen from Eq.~\ref{eq:projS}. Dominant singlet pairing mainly occurs 
on bonds between sites within the unit-cell, with a weaker pairing on bonds 
connecting the diamond plaquettes. The largest pairing amplitude at any $J_1$ appears
for densities $\bar{n} \approx 0.125$ which coincides with the 2D van Hove singularity
in the density of states seen in Fig.~\ref{fig:Band_Structure}.
The structure of the pair amplitude on the unit cell is shown in Fig.~\ref{fig:pattern}(a), 
and the momentum dependent spectral gap at the chemical potential is shown in Fig.~\ref{fig:spectra}(a).

\medskip

(ii) {\bf AFM order:} In both cases, we find a weak coupling instability at half-filling
($\bar{n}=0.5$) towards N\'eel AFM driven by nesting of the Fermi surfaces. We have
confirmed this also within a random phase approximation (RPA) calculation discussed in the
Appendix \ref{appendix:magnetic}. \textcolor{black}{For the case $J_2 > 0$, this Stoner-type AFM order 
persists to strong coupling, 
much like in the one-band Hubbard model, leading to an AFM insulator.} By contrast, for $J_2 < 0$, 
this Stoner AFM is lost as we move towards strong coupling.
Fig.~\ref{fig:pattern} (e) 
depicts the magnetization in this AFM state.

\medskip

(iii) {\bf Strong {\alm} order:} At half-filling, for $J_2 < 0$, going to strong coupling
leads to {\alm} order with a large ordered moment.
\textcolor{black}{This state remains metallic for {\alm} Weiss fields $h < 2.0 t_1$. At this critical value, the Fermi surface gaps out and the system transitions to an insulating {\alm} phase. This insulating phase appears for $J_1 > 4.0 t_1$,
and is driven by local moments coupled through $J_1 > 0$ and $J_2 < 0$, as can be qualitatively
understood through an effective spin-only model for the Mott insulator.}
Fig.~\ref{fig:pattern} (d)
shows the structure of the magnetization in this A$\ell$M state.

\medskip

(iv) \textcolor{black}{{\bf Itinerant {\alm}/{SDW} order:} Remarkably, for 
$J_2 < 0$, we find a second window of {\alm} order
at weak coupling, for densities $0.26 \lesssim \bar{n} \lesssim 0.3$,
featuring a small ordered moment. This {\it itinerant} {\alm} phase can be understood 
as being driven by the quasi-1D
vHS seen in Fig.~\ref{fig:Band_Structure} corresponding to a low filling of the $X,Y$
bands. In this regime, it is well known that the divergent DOS leads to a 
strongly enhanced ferromagnetic Stoner susceptibility for the quasi-one-dimensional X and Y bands. 
The weak {\alm} phase then results 
from the effective AFM Hund's coupling seen in Eq.~\ref{eq:projXY}
between weakly ferromagnetic X and Y chains in the projected two-orbital model. The itinerant
{\alm} state has the same symmetry as the strong coupling
{\alm}, except for being metallic and featuring a highly reduced moment. Such an itinerant {\alm} order would also
be stabilized if we set $J_2=0$ but instead include intra-cluster or inter-cluster further neighbor
repulsive interactions, so it could arise purely from an extended Hubbard model on the square-octagon lattice.
We expect that this van Hove mechanism of {\alm} order might also arise naturally in realistic models and materials
with quasi-1D bands.}

For \(J_2>0\), we find a weak SDW ordering 
in the same density regime; however, the ordering wave-vector 
is \(\boq \approx \boldsymbol{M}/2\), with a \(d\)-type ordering within the unit cell 
(similar to the \alm), so we label it as $d-$SDW.
This ordering wavevector corresponds to the approximate nesting of the 
Fermi surface at these densities; this instability is bolstered by interactions
(see Appendix \ref{appendix:magnetic} for results from RPA calculation).

\medskip

(v) {\bf Uniform $d$-wave SC:} Both models, with weak $J_2 < 0$ and $J_2 > 0$, show large
regimes of uniform singlet $d$-wave superconductivity for $0.25 < \bar{n} < 0.5$. This is primarily
driven by the formation of singlet Cooper pairs on bonds with the unit cell plaquettes, driven by AFM 
$J_1 > 0$, which can then delocalize across the lattice. In the projected model, this phase
can be simply viewed as the local AFM Hund's coupling leading to a nonzero on-site order parameter 
$\langle X^\dagger Y^\dagger \rangle$. Since $C_4$ rotation leads to $X \to Y$ and $Y \to -X$, this
local order parameter changes sign in accordance with $d$-wave symmetry; more precisely, this is a 
SC with a $d_{xy}$ order parameter.
In several instances, we have found that real space solutions
of the mean field equations starting from random initial conditions result in $d$-wave droplets 
which are not fully phase-locked across the lattice, suggestive of a not-very-strong superfluid 
stiffness; in future work, we will address the superfluid stiffness in these regimes. However, 
in all these cases, we have checked that the uniform $d$-wave solution has a slightly lower energy than
these phase-random $d$-wave droplet solutions. Fig.~\ref{fig:pattern}(b) and  ~\ref{fig:spectra}(b)
show the pair amplitude in real space, and the momentum
dependent spectral gap at the chemical potential.

\medskip

(vi) {\bf $d$-wave PDW order:} Interestingly, for $J_1 \lesssim 1.0$ and for $ 0.38 \lesssim n \lesssim 0.48$, 
we find regimes of $(\pi,\pi)$ $d$-wave PDW order in our inhomogeneous 
mean field theory on system sizes upto
$6\times 6$ unit cells. In this $d$-PDW, electrons form singlet $d$-wave
pairs within the unit cell but the pairing amplitude changes sign in the adjacent unit cell.
Fig.~\ref{fig:pattern}(c)
shows the structure of the pair amplitude on the unit cell, and Fig.~\ref{fig:spectra}(c) shows 
the resulting momentum dependent spectral gap in the full Brillouin zone. The
RPA pairing susceptibility in the normal state shows significant
peaks at $(0,0)$ and $(\pi,\pi)$ (see Appendix \ref{appendix:pairing}); however,
the most divergent low temperature susceptibility is at $(0,0)$, suggesting that this
finite size
$d$-wave PDW order should disappear with increasing system size as we have confirmed using momentum space
mean field theory. However, the fact that the PDW state is favored on
small system sizes suggests that slightly tuning the interactions, perhaps by including further neighbor
interactions could favor this as the true ground state, as suggested
in recent work on other models \cite{pdw_ticea2024}.

\medskip

\begin{figure*}[!t]
\centering
\includegraphics[width=0.24\textwidth,height=0.24\textwidth]{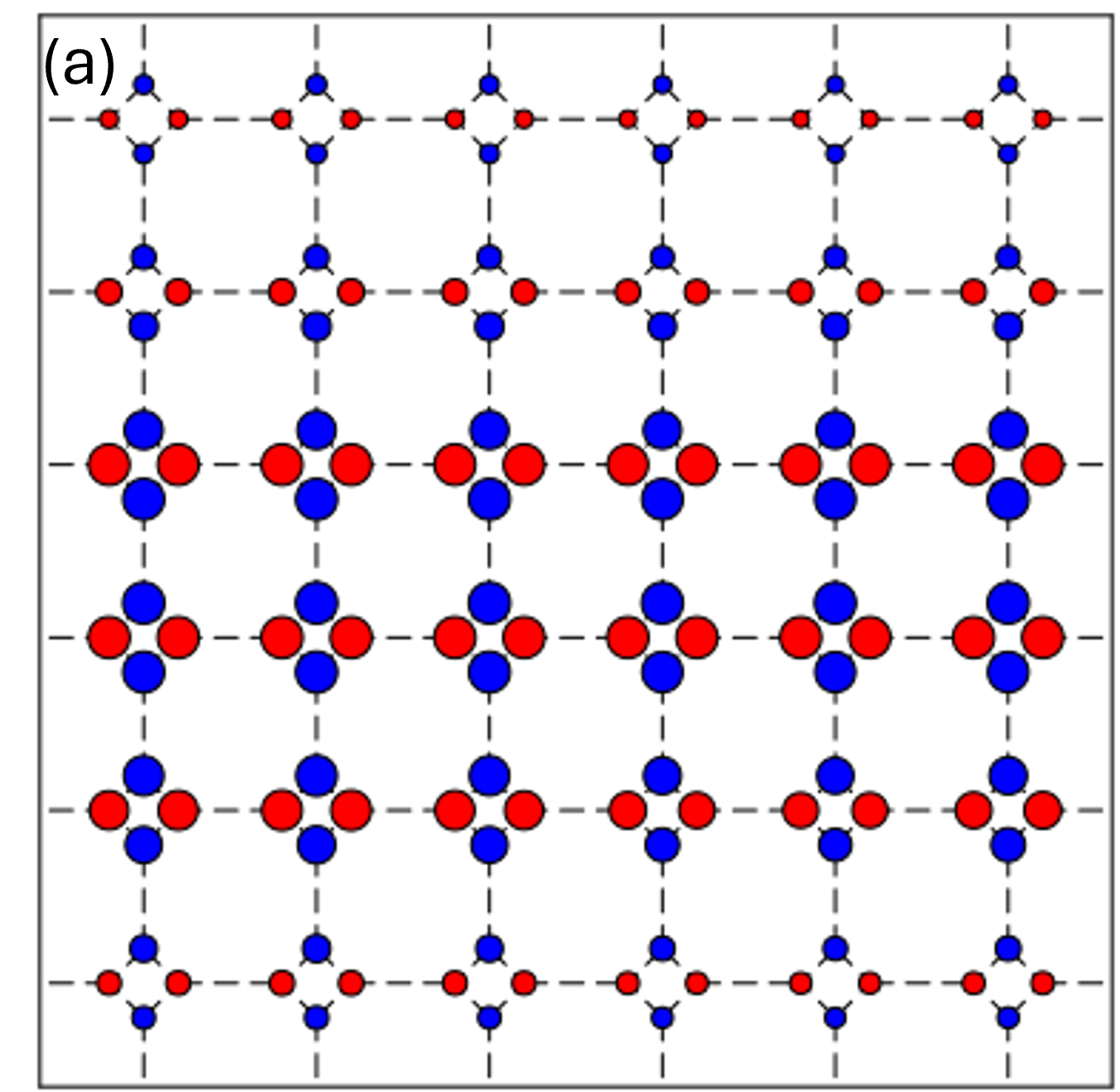}
\includegraphics[width=0.24\textwidth,height=0.24\textwidth]{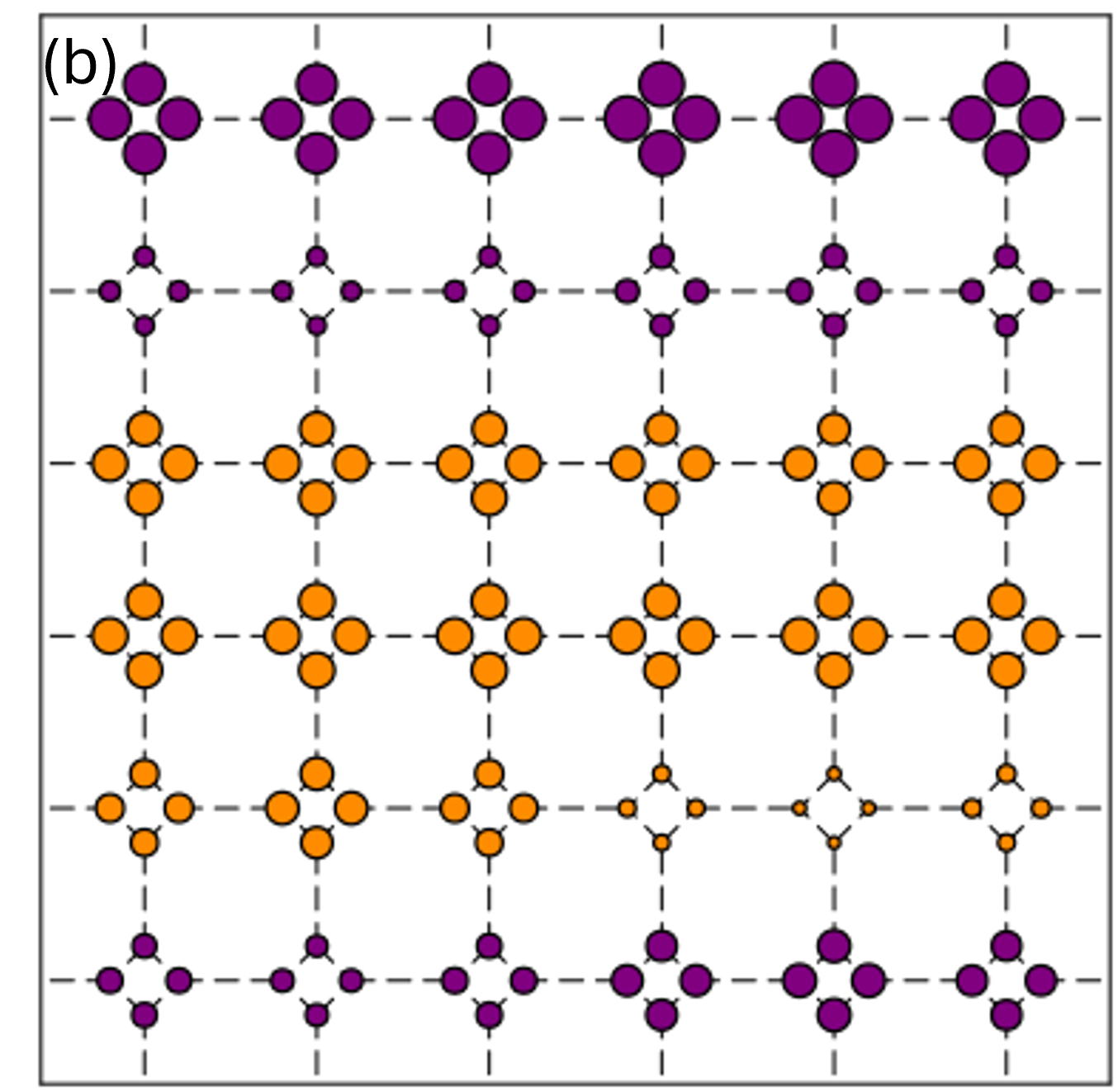}
\includegraphics[width=0.24\textwidth,height=0.24\textwidth]{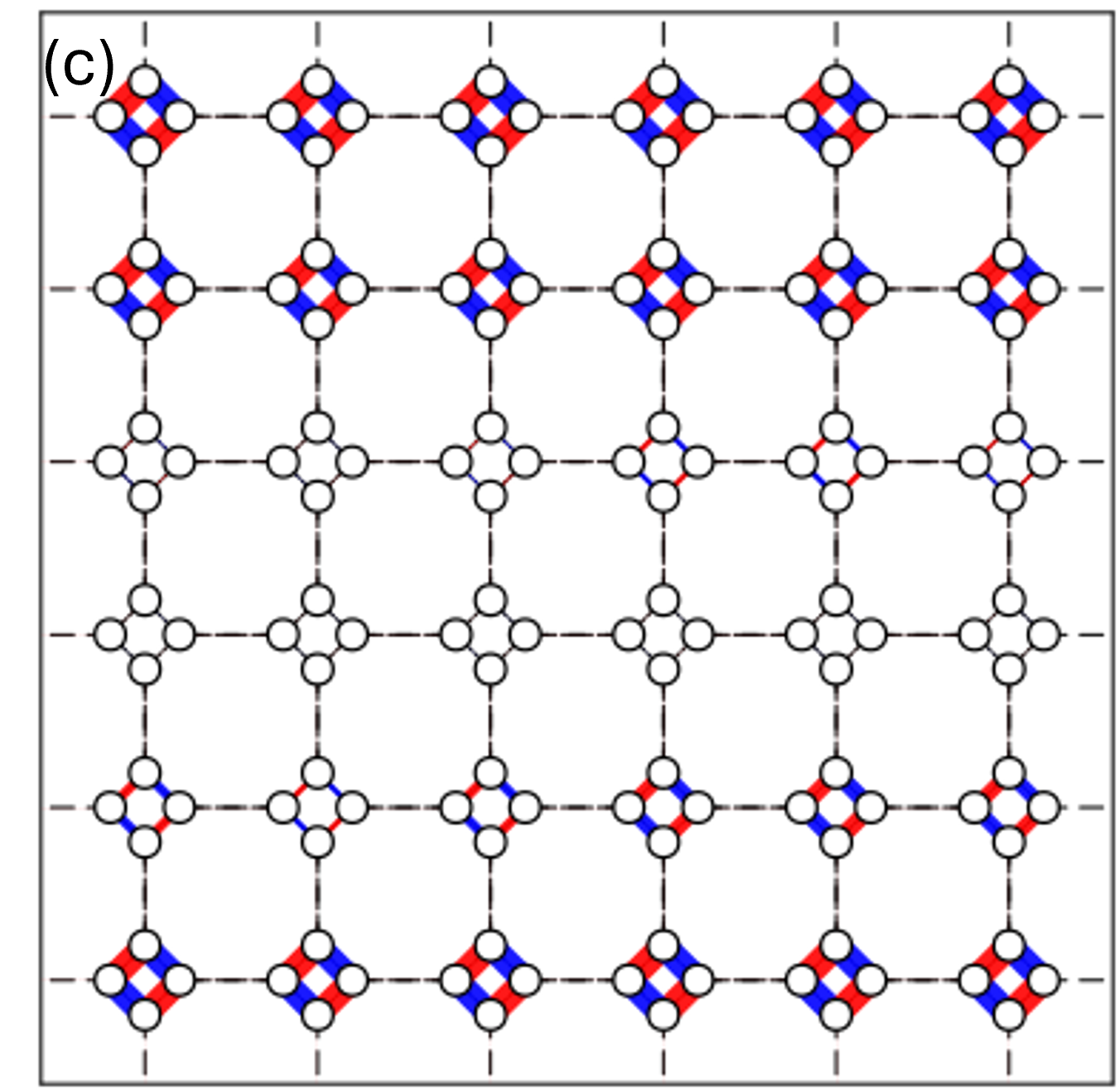}
\includegraphics[width=0.24\textwidth,height=0.24\textwidth]{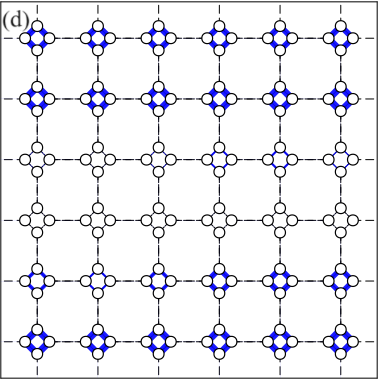}
\caption{Example of converged real space mean field solution at $\bar{n}=0.46$, $J_2=-0.5 J_1$, and $J_1=3.6$, showing
phase separated state. (a) Inhomogeneous magnetic order of 
{\alm} type with strong and weak {\alm} regions. The weak {\alm} order is induced by proximity to the strong {\alm} region. black marks an on-site spin-up while red marks spin-down.
(b) Density modulations which track the {\alm} modulations
with strong {\alm} order corresponding to density $\bar{n} \approx 0.5$ and weak {\alm} regions having density $\bar{n} < 0.46$. 
Orange marks density higher than the mean filling while purple denotes the opposite.
(c) spatially inhomogeneous singlet pairing, and (d) spatially inhomogeneous triplet pairing. The 
mixture of $d$-wave singlet and $s$-wave triplet pairing in regions of low {\alm} order is also driven by proximity to the {\alm} domain,
similar in spirit to the work by Brekke, {\it et al} \cite{AM_Brekke2023TwodimensionalAS}. black and red denote opposite phases in the pairing which give it the \(d\)-wave nature. Appendix \ref{appendix:symmetries} presents an analysis of the different pairing phases observed in the phase diagram.} 

\label{fig:phasesep}
\end{figure*}

(vii) {\bf Coexistence / Phase separation:} Finally, in the strong coupling regime and for small doping away
from half-filling, we find a regime where superconductivity and AFM or {\alm} coexist in uniform
mean field theory as can be seen from the momentum space phase diagram in Appendix \ref{appendix:kSpace}. 
In this coexistence phase, the {\alm} order mixes the real space
singlet $d$-wave and triplet $s$-wave pairing with $S_z=0$ on the diamond unit cell, which is allowed by
the presence of multiple orbitals.
Based on the symmetry analysis in Appendix \ref{appendix:symmetries} of possible pairing states, and given the
proximity of the coexistence phase to the uniform $d$-wave singlet phase, we find that
the superconductor coexisting with the {\alm} phase is 
a superposition of \(d-\)wave singlet with \(s-\)wave triplet within the unit cell.
These two states mix because \(\mathcal{C}_4\) and \(\mathcal{T}\) are broken by the {\alm}, 
but \(\mathcal{C}_4\otimes\mathcal{T}\) is preserved under which both these pairing states are odd as shown in 
Table.\ref{table:symmetries}.
Allowing for a spatially inhomogeneous state, these coexisting orders
phase separate, with
the magnetically ordered regimes having density $\bar{n} \approx 0.5$ and forming {\alm} droplets
while the lower density regimes have $d$-wave pairing. In this regime, a uniform ansatz favors coexistence of
AFM / {\alm} order with superconductivity. This coexistence leads to mixed singlet-triplet pairing within 
the unit cell; such {\alm}-induced triplet pairing has been explored in recent work discussing {\alm} fluctuation
induced pairing \cite{AM_Brekke2023TwodimensionalAS,tripletAM_Brataas2024, tripletAM_Zhang2024,maeland2024manybody}. 
Incorporating longer-range repulsion between doped charges could lead to organization of
these phase separated regions into stripe or checkerboard states as has also been found in models for
the cuprate superconductors. Fig.~\ref{fig:phasesep} 
shows the typical structure of the density, magnetization, and pairing amplitude in such a phase separated
state.

\section{Summary and discussion}
Motivated by exploring $t$-$J$ models which capture {\alm} order and to study the possible effects of doping and pressure-tuning of bandwidth, we have
discussed a mean field theory of a minimal model on the square-octagon lattice. This is shown to have a phase diagram which includes multiple
types of {\alm} order - \textcolor{black}{a strongly ordered {\alm} phase with large moment at strong coupling and half-filling,
and a weakly ordered itinerant {\alm} metallic phase driven by quasi-1D van Hove singularities.}
The latter suggests a possible alternative design route to realizing weak \textcolor{black}{itinerant {\alm} order} in materials. 
In addition, our phase diagram features multiple
types of superconducting orders including $s$-wave, $d$-wave, and possible $d$-wave PDW order. For $J_2 >0$ coupling between unit cells, this leads
to locally {\alm}-type configuration which is modulated at an incommensurate wavevector $Q$ which we have dubbed $d$-SDW. 
We discussed the spatial structure of the various
broken  symmetry states and  presented aspects of the spectral gap features in these phases. We are currently exploring if 
incorporating additional interactions in this model could stabilize stable PDW order. It would be interesting to also explore if such a rich set of
orders are generic in other {\alm} models with orbital and spin order which have been explored in recent work \cite{leeb2023spontaneous,das2023realizing}
which would provide further
motivation for experimentalists to search for superconductivity in doped or pressurized {\alm} materials. Finally, incorporating weak 
spin-orbit coupling effects in these $tJ$ models might be a route to interesting topological superconducting phases and would be a useful direction for
future research.

\acknowledgments

This research was funded by the Natural Sciences and Engineering Research Council (NSERC) of Canada.
SV was supported by a scholarship from the Fonds de recherche du Qu\'ebec - Nature et Technologies (FRQNT, Quebec).
Numerical computations were performed on the Niagara supercomputer at the SciNet HPC Consortium and the Digital Research Alliance of Canada.  The Julia
codes for mean-field simulations and tight-binding analysis are available online at \href{https://github.com/Anjishnubose/TightBindingToolkit.jl}{TightBindingToolkit.jl} and \href{https://github.com/Anjishnubose/MeanFieldToolkit.jl}{MeanFieldToolkit.jl}. The bare susceptibility results were done in Julia as well as using TRIQS \cite{Parcollet_2015}

\bibliography{Altermagnetism}

\appendix 
\onecolumngrid
\clearpage

\section{Susceptibilities}
To get an idea of possible instabilities, one can calculate the susceptibility of these multi-orbital or multi-sublattice model towards them. The formulation is a bit more involved but essentially the same as the simple single-band case. Furthermore, to leading order, one can also look at the effects of the interaction under RPA type approximation. To begin by looking at bare susceptibilities, let us define the generalized order parameter (OP) fields  corresponding to our orderings as
\begin{itemize}
    \item \textbf{Magnetic order} : Define the local magnetization OP fields as \(M_{\mathbf{r}, i}^{a}(\t) = (1/2)c^{\dagger}_{\mathbf{r}, i, \a}(\t) \s^{a}_{\a\b} c_{\mathbf{r}, i, \b}(\t)\), where \(\mathbf{r}\) refers to the unit-cell position, \(i\) refers to every degree of freedom other than spin (such as sublattice or orbital), \(a=x,y,z\) refers to the spin-ordering direction, and \(\a,\b\) refers to the fermion spin. In momentum space the corresponding vertex looks like
    \begin{equation}
        \label{magnetic vertex}
        M_{\boq, i}^{a}(i\W) = \s^{a}_{\a\b}\int \frac{d^2k}{(2\p)^2}\int \frac{d\w}{(2\p)} c^{\dagger}_{\boq+\bok, i, \a}(i\W+i\w) c_{\bok, i, \b}(i\w)\,,
    \end{equation}
    where \(\boq\) is the exchange momentum and \(\W\) is the exchange energy. 
    \item \textbf{Pairing order} : Now pairing is more complicated since one can have non-local pairing as well. To that end, a generalized pairing OP field will look like \(\D_{\mathbf{r+\d/2}}^{a, ij}(\t) = (1/2)c^{\dagger}_{\mathbf{r}, i, \a}(\t)\s^{a}_{\a\b}c^{\dagger}_{\mathbf{r+\d},j,\b}(\t)\) corresponding to a non-local pairing when \(\d\neq 0\). Such OP fields generically live on the bonds of the lattice as opposed to the magnetic ordering field living on-site. Again, in Fourier space, the vertex (which is now momentum dependent) looks like
    \begin{equation}
        \label{pairing vertex}
        \D_{\boq, ij}^{a}(\bod ; i\W) = \s^{a}_{\a\b}\int \frac{d^2k}{(2\p)^2}\int \frac{d\w}{(2\p)} c^{\dagger}_{-\bok + \boq/2, i, \a}(-i\w + i\W/2) c^{\dagger}_{\bok + \boq/2, j, \b}(i\w + i\W/2) e^{i\bod\vdot\bok}\,,
    \end{equation}
    where \(\boq\) is the center of mass momentum and \(\W\) is the center of mass energy.
\end{itemize}

Using these OP fields, we can calculate the bare susceptibility of the model as shown in the subsequent sections. To get ordering tendencies of the model, we have to diagonalize the zero-energy response matrix at all momenta. The eigenvector with the largest eigenvalue corresponds to possible orderings with the ordering vector being the momentum at which this maximum eigenvalue occurs. One can repeat this exercise after performing RPA, which can affect possible orderings in the model as interaction strength is slowly increased. Encountering a diverging eigenvalue of the response corresponds to a phase transition into a symmetry broken state, which should match qualitatively with mean-field results. Alternatively, one can also tune the temperature at strong coupling regime to extract information about phases beyond the first instability encountered at zero temperature when tuning the interaction.

\subsection{Magnetic channel}\label{appendix:magnetic}

The bare susceptibility, \(\c_{0, ij}^{ab}(\mathbf{Q}, i\Omega) = \expval{M_{\mathbf{Q},i}^{a}(i\Omega)M_{\mathbf{Q}, j}^{b, \dagger}(i\Omega)}\) is equivalent to the usual spin-response function. Diagrammatically, it corresponds to a generalized bubble diagram. For spin-rotation symmetric systems, the connected piece of the diagram looks like
\begin{equation}
    \label{magnetic suscep 1}
    \chi_{0, ij}^{ab}(\mathbf{Q}, i\Omega) = -\frac{\d^{ab}}{4}\int \frac{d^2k}{(2\p)^2}\int \frac{d\w}{(2\p)} G_{ij}(\mathbf{Q}+\mathbf{k}, i\Omega+i\omega)G_{ji}(\mathbf{k}, i\w)\,,
\end{equation}
where \(G_{ij}^{\a\b}(\mathbf{k}, i\w) = \d^{\a\b}G_{ij}(\mathbf{k}, i\w) = \expval{c_{i,\a}(\mathbf{k}, i\w) c^{\dagger}_{j, \b}(\mathbf{k}, i\w)}\) are the bare Green's functions calculated in mean-field theory. In terms of the quasi-particle dispersion, \(\e_n(\mathbf{k})\), and their corresponding wavefunctions \(U_{ij}(\mathbf{k})\), these Green's functions can be expressed as
\begin{equation}
    \label{greens function wavefunctions}
    G_{ij}(\mathbf{k}, i\w) = \sum_{n}U_{in}(\mathbf{k})\frac{1}{i\w - \e_n(\mathbf{k}) + \mu}U^{\dagger}_{nj}(\mathbf{k})\,.
\end{equation}
Substituting this expression back into \eqref{magnetic suscep 1}, performing the Matsubara sum by hand, and analytically continuing the resulting expression \(i\Omega\rightarrow\Omega+i\eta\), we get
\begin{equation}
    \label{magnetic suscep}
    \chi_{0, ij}^{ab}(\mathbf{Q}, \Omega) = \frac{\d^{ab}}{4}\sum_{n,m}\int \frac{d^2k}{(2\p)^2} U_{jn}(\mathbf{Q}+\mathbf{k})U_{ni}^{\dagger}(\mathbf{Q}+\mathbf{k})U_{im}(\mathbf{k})U_{mj}^{\dagger}(\mathbf{k}) \frac{n_F(\e_{n}(\mathbf{Q}+\mathbf{k})-\mu) - n_F(\e_m(\mathbf{k})-\mu)}{\Omega + i\eta - (\e_{n}(\mathbf{k}+\mathbf{Q}) - \e_m(\mathbf{k}))}\,,
\end{equation}
where \(n_F\) is the Fermi distribution function.\par

To perform RPA, let us focus on analyzing Heisenberg type interactions and decompose it in terms of the magnetic OP fields. We have a generalized Heisenberg type interaction on a lattice as
\begin{equation}
    \label{general heisenberg}
    \mh_{int} = \frac{1}{4}\sum_{\bod}\sum_{\bor, \t}J_{ij}(\bod) c^{\dagger}_{\bor, i, \a}(\t) \s^{a}_{\a\b} c_{\bor, i, \b}(\t)c^{\dagger}_{\bor+\bod, j, \m}(\t) \s^{a}_{\m\n} c_{\mathbf{r}+\bod, j, \n}(\t)\,.
\end{equation}
Using the definition of the magnetic OP fields, we immediately see that
\begin{equation}
    \label{heisenberg decomposition magnetic}
    \begin{split}
        \mh_{int} &= \sum_{\bod}\sum_{\bor, \t}J_{ij}(\bod)M_{\bor, i}^{a}(\t) M^{a, \dagger}_{\bor + \bod, j}(\t)\\
        &= \sum_{\boq, \Omega}M_{\boq, i}^{a}(\Omega)\left[\d^{ab}\sum_{\bod}J_{ij}(\bod)e^{-i \boq\vdot\bod}\right] M^{b, \dagger}_{\boq, j}(\Omega)\,.
    \end{split}
\end{equation}
Hence, under RPA, the susceptibility will have the form
\begin{equation}
    \label{magnetic RPA}
    \chi_{ij}^{ab}(\boq, \W) = \left[\left(\mathds{1} - \chi_{0}(\boq, \Omega)\vdot \Tilde{J}(\boq)\right)^{-1}\vdot \chi_{0}(\boq, \Omega)\right]_{ij}^{ab}\,,
\end{equation}
where the effective interaction matrix is \( \Tilde{J}_{ij}^{ab}(\boq) = \d^{ab}\sum_{\bod}J_{ij}(\bod)e^{-i \boq\vdot\bod}\), and \(\vdot\) denotes matrix multiplication in the spin-direction index \(a\) and sublattice/orbital index \(i\) .\par

\begin{figure*}
    \centering
    \includegraphics[width=0.4\textwidth]{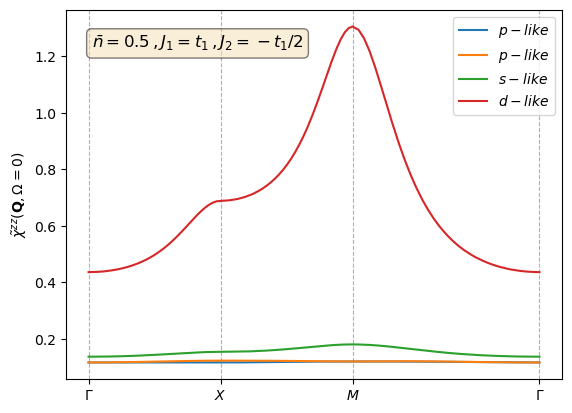}
    \includegraphics[width=0.4\textwidth]{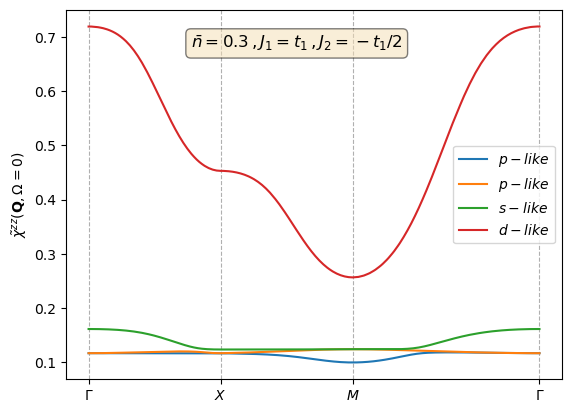}
    \caption{The magnetic susceptibilities at zero-energy at a fixed low temperature of \(T = 0.1 t_1\) after performing RPA at \(J_1 = t_1, J_2 = -0.5t_1\),plotted along a high-symmetry path in the Brillouin zone, at two different fillings. The labels are the same as in Fig.\ref{fig:mag_suscep} (a) At half-filling,\(\bar{n} = 0.5\), the ordering at \(Q = M\) with a $d$-like eigenvector corresponding to the usual AFM ordering. (b) At some generic filling, \(\bar{n}=0.3\), the RPA susceptibility is peaked at \(Q=\boldsymbol{\Gamma}\), again with the \(d\) type eigenvector. This corresponds to a {\alm} type ordering. }
    \label{fig:mag_suscep_RPA_FM}
\end{figure*}

\begin{figure*}
    \centering
    \includegraphics[width=0.4\textwidth]{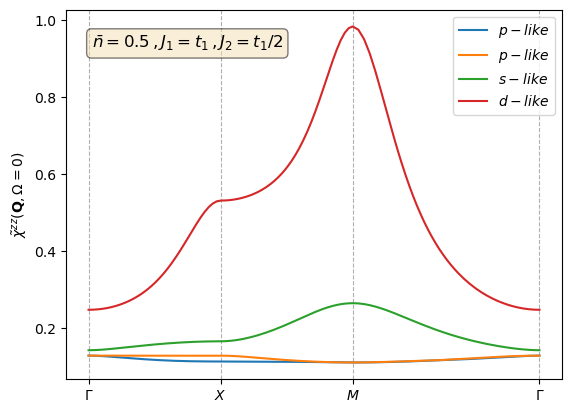}
    \includegraphics[width=0.4\textwidth]{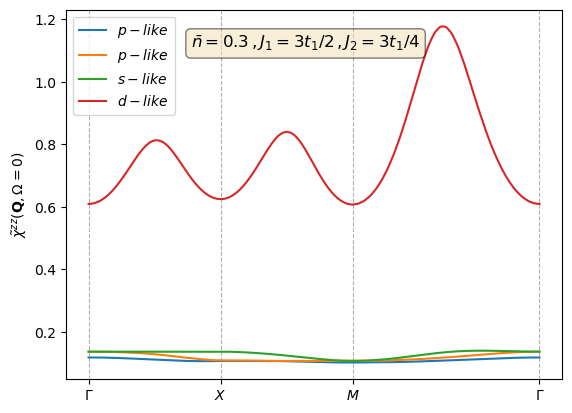}
    \caption{The magnetic susceptibilities with \(J_2/J_1>0\) at zero-energy at a fixed low temperature of \(T = 0.1 t_1\) after performing RPA, plotted along a high-symmetry path in the Brillouin zone, at two different fillings. The labels are the same as in Fig.\ref{fig:mag_suscep} (a) At half-filling,\(\bar{n} = 0.5\), the ordering at \(Q = M\) with a \(d-like\) eigenvector corresponding to the usual AFM ordering. (b) At \(\Bar{n}=0.3\) close to the Van-hove singularities, the ordering is at \(Q~M/2\) with a d-type eigenvector. }
    \label{fig:mag_suscep_RPA_AFM}
\end{figure*}

\begin{figure*}
    \centering
    \includegraphics[width=0.4\textwidth]{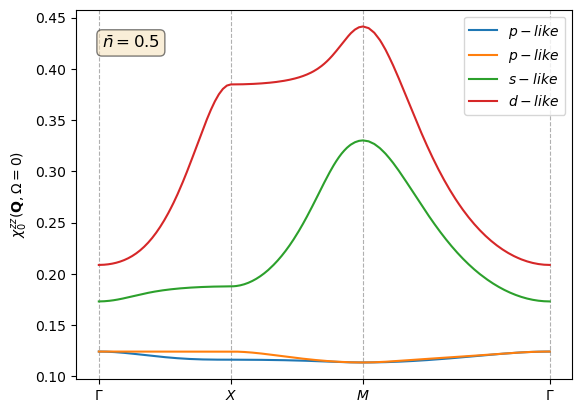}
    \includegraphics[width=0.4\textwidth]{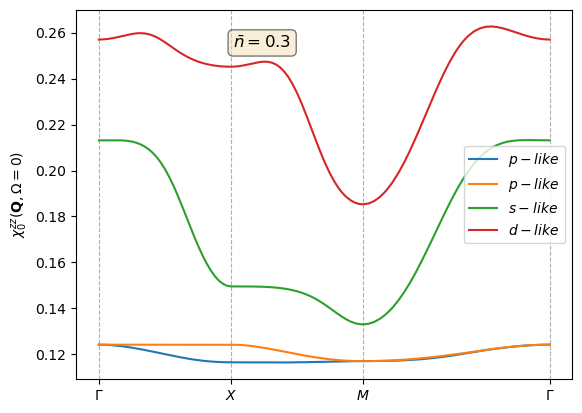}
    \caption{The bare magnetic susceptibilities at zero-energy at a fixed low temperature of \(T = 0.1 t_1\) plotted along a high-symmetry path in the Brillouin zone, at two different fillings. The labels describe the eigenvectors of the susceptibility matrix such that \(s-like\) corresponds to all spins pointing parallel to each other in the unit cell, \(d-like\) corresponds to the spins alternating within the unit cell and so on. (a) At half-filling,\(\bar{n} = 0.5\), where we find the ordering at \(Q = M\) with a \(d-like\) eigenvector. This just corresponds to a usual AFM ordering. (b) At some generic filling, \(\bar{n}=0.3\), we find more complicated momentum-dependent behaviour which is peaked \emph{near} \(Q=\boldsymbol{\Gamma}\), again with the \(d\) type eigenvector. }
    \label{fig:mag_suscep}
\end{figure*}

For our model, with \(J_2/J_1<0\) shown in Fig.\ref{fig:mag_suscep_RPA_FM}, we find that at exactly half-filling \(\bar{n}=0.5\), there is an instability towards AFM order which can be seen by the response peaking near the \(Q = \boldsymbol{M}=(\pi,\pi)\) point in the Brillouin zone. However, at every other generic filling, the system prefers {\alm} ordering, peaked at the \(\boldsymbol{\G}\) point. Whereas, for \(J_2/J_1>0\) as shown in Fig.\ref{fig:mag_suscep_RPA_AFM}, the instability towards AFM at half-filling is still there, but the instability at lower filling near the van-Hove singularities is now with \(Q = \boldsymbol{M}/2\). This comes up due to Fermi-surface nesting around that filling, as can be seen even in the bare susceptibilities in Fig.\ref{fig:mag_suscep}.

\subsection{Pairing channel}\label{appendix:pairing}

Let us look at the bare pairing susceptibility by repeating the steps as in the magnetic case (with a lot more indices to keep track of!). We want to calculate \(\G_{0, ijkl}^{ab}(\bod_1, \bod_2 ; \boq, i\W) = \expval{\D_{\boq, ij}^{a}(\bod_1 ; i\W) \D_{\boq, kl}^{b, \dagger}(\bod_2, i\W)}\). Working this bubble out in all its gory details gives us
\begin{equation}
    \label{pairing bubble}
    \begin{split}
        \G_{0, ijkl}^{ab}(\bod_1, \bod_2 ; \boq, i\W) = \frac{1}{4}\s^{a}_{\a\b}\s^{b}_{\m\n}\int \frac{d^2k}{(2\p)^2}\int \frac{d\w}{(2\p)} \bigg\{ & e^{i\bok\vdot(\bod_1-\bod_2)}G_{ik}^{\a\d}(-\bok+\boq/2, -i\w+i\W/2)G_{jl}^{\b\g}(\bok+\boq/2, i\w+i\W/2)\\
        -& e^{i\bok\vdot(\bod_1+\bod_2)}G_{il}^{\a\m}(-\bok+\boq/2, -i\w+i\W/2)G_{jk}^{\b\n}(\bok+\boq/2, i\w+i\W/2)\bigg\}\,.
    \end{split}
\end{equation}
We can simplify the above expression when we use spin-rotation symmetry which tells us that the pairing response can be categorized into singlet and triplet pairings. Defining a shorthand for one of the two contributions to the integral as
\begin{equation}
    \label{integral shorthand}
    \mi_{ijkl}(\bod ; \boq, i\W) \equiv \int \frac{d^2k}{(2\p)^2}\int \frac{d\w}{(2\p)}  e^{i\bok\vdot\bod}G_{ik}(-\bok+\boq/2, -i\w+i\W/2)G_{jl}(\bok+\boq/2, i\w+i\W/2)\,,
\end{equation}
we get that
\begin{equation}
    \G_{0, ijkl}^{ab}(\bod_1, \bod_2 ; \boq, i\W) = 
    \begin{cases}
    \frac{1}{2}\mi_{ijkl}(\bod_1-\bod_2 ; \boq, i\W) - \frac{1}{2}\mi_{ijlk}(\bod_1+\bod_2 ; \boq, i\W)\,,\:\:\:a=b\in\{0, x, z\} : \text{ triplet}\,,\\
    \frac{1}{2}\mi_{ijkl}(\bod_1-\bod_2 ; \boq, i\W) + \frac{1}{2}\mi_{ijlk}(\bod_1+\bod_2 ; \boq, i\W)\,,\:\:\:a=b\in\{y\} : \text{ singlet}\,,\\
    0 \,,\:\:\: a\neq b\,,
    \end{cases}
\end{equation}
where \(\s^{0} = \mathds{1}_{2\times2}\), the identity matrix. Focusing on \(\mi_{ijkl}\), we can repeat the steps as before \textit{i.e.} write everything using the band dispersion and wavefunctions, performing the Matsubara sum by hand, and analytically continuing the resultant expression. We end up with
\begin{equation}
    \label{integral simplified}
    \begin{split}
        \mi_{ijkl}(\bod ; \boq, \W) = \sum_{n, m}\int \frac{d^2k}{(2\p)^2} & U_{kn}(-\bok+\boq/2)U^{\dagger}_{ni}(-\bok+\boq/2) U_{lm}(\bok+\boq/2)U^{\dagger}_{mj}(\bok+\boq/2)\\
        \times & e^{i\bok\vdot\bod}\cdot\left(\frac{n_F(\e_{n}(-\bok+\boq/2) - \mu) + n_F(\e_{m}(\bok+\boq/2) - \mu) - 1}{\W + i\eta - \e_{n}(-\bok+\boq/2) - \e_{m}(\bok+\boq/2) + 2\mu}\right)\,.
    \end{split}
\end{equation}

Now, for RPA in this channel, we will have to redo the decomposition of the interaction in terms of the pairing OP fields. We again start with \eqref{general heisenberg} and rearrange the terms as
\begin{equation}
    \label{heisenberg pairing}
    \mh_{int} = \frac{1}{4}\s^{a}_{\a\b}\s^{a}_{\m\n}\sum_{\bod}\sum_{\bor, \t}J_{ij}(\bod) c^{\dagger}_{\bor, i, \a}(\t) c^{\dagger}_{\bor+\bod, j, \m}(\t)  c_{\mathbf{r}+\bod, j, \n}(\t) c_{\bor, i, \b}(\t)\,.
\end{equation}

\begin{figure*}[ht]
    \centering
    \includegraphics[width=0.5\textwidth]{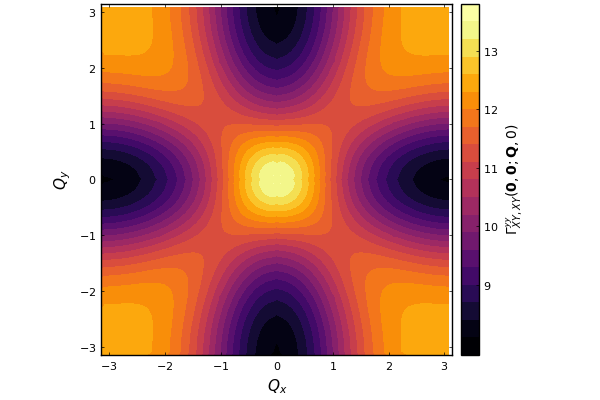}
    \caption{The normal state RPA singlet pairing susceptibility with $t_1=1$, $t_2=0.5 t_1$, $J_1=0.5 t_1$ and $J_2 = -0.5 J_1$ 
    at $\Omega=0$ and a fixed low temperature of \(T = 0.13 t_1\), plotted 
    on the Brillouin zone for the reduced two-orbital model. The largest eigenvalue comes from the \(X-Y\) singlet pairing on-site which corresponds to a \(d\)-wave pairing in the original model. }
    \label{fig:pair_suscep RPA}
\end{figure*}

We see that we cannot simply substitute the form of our pairing OP fields since the spin-indices are not as we want them in \eqref{heisenberg pairing}. However, since the Heisenberg interaction is spin-rotation symmetric, we can again decompose it in terms of singlet and triplets. It turns out that the interaction then looks like
\begin{equation}
    \label{heisenberg pairing singlet-triplet}
    \begin{split}
        \mh_{int} &= \frac{1}{4}\sum_{\bod}\sum_{\bor, \t}J_{ij}^{ab}(\bod) \left(c^{\dagger}_{\bor, i, \a}(\t)\s^{a}_{\a\m} c^{\dagger}_{\bor+\bod, j, \m}(\t)\right) \left( c_{\mathbf{r}+\bod, j, \n}(\t) \s^{b}_{\n\b}c_{\bor, i, \b}(\t)\right)\\
        &= \sum_{\bod}\sum_{\bor, \t}J_{ij}^{ab}(\bod) \D_{\bor+\bod/2}^{a, ij}(\t) \D_{\bor+\bod/2}^{b, ij, \dagger}(\t)\,,\\
        &= \sum_{\boq, \W}\D_{\boq, ij}^{a}(\bod_1 ; \W) \left[\d^{(ij), (kl)}\d_{\bod_1, \bod_2}\sum_{\bod_1}J_{ij}^{ab}(\bod_1) \right] \D_{\boq, kl}^{b, \dagger}(\bod_2 ; \W)\,,
    \end{split}
\end{equation}
where the interaction in the singlet-triplet pairing basis looks like
\begin{equation}
    \label{interaction spin resolved}
    J_{ij}^{ab}(\bod) = 
    \begin{cases}
        \frac{1}{2}J_{ij}(\bod)\,,\:\:\:a=b\in\{0, x, z\} : \text{ triplet}\,,\\
        -\frac{3}{2} J_{ij}(\bod)\,,\:\:\:a=b\in\{y\} : \text{ singlet}\,,\\
        0\,,\:\:\:a\neq b\,.
    \end{cases}
\end{equation}
Note that the interaction is completely \emph{momentum independent} in the center of mass basis, and diagonal in the pairing OP fields! The total RPA susceptibility takes on a similar form as \eqref{magnetic RPA} as
\begin{equation}
    \label{pairing RPA}
    \G_{ijkl}^{ab}(\bod_1, \bod_2 ; \boq, i\W) = \left[\left(\mathds{1} - \G_{0}(\boq, i\W) \vdot \mj(\boq)\right)^{-1}\vdot \G_{0}(\boq, i\W) \right]_{ijkl}^{ab}(\bod_1, \bod_2)\,,
\end{equation}
where  \(\mj_{ij, kl}^{ab ; \bod_1\bod_2}(\boq) = \d^{(ij), (kl)}\d_{\bod_1, \bod_2}\sum_{\bod}J_{ij}^{ab}(\bod_1)\), and \(\vdot\) now denotes matrix multiplication in all possible index degree of freedom including spin \(a\), bond displacement \(\bod\), and pairing orbital/sublattices \(ij\).\par

For simplicity, we chose to use the reduced two-orbital model instead of the full square-octagon model, and focus on singlet pairing susceptibility. The pairing OP fields we choose to work with are the ones which can be generated through our interaction, namely \(X-Y\) pairing on site \(\D_{\boq, XY}^{y}(\mathbf{0} ; 0)\), \(X-X\) pairing along the \(x\)-bonds \(\D_{\boq, XX}^{y}(\hat{x} ; 0)\), and \(Y-Y\) pairing along the \(y\)-bonds \(\D_{\boq, YY}^{y}(\hat{y} ; 0)\). Note that since \(\D_{\boq, ij}^{a}(\bod ; i\W) = \D_{\boq, ji}^{a}(-\bod ; i\W)\), we do not need to track \(\bod = -\hat{x}\) and \(\bod = -\hat{y}\) separately.\par

We find that the largest eigenvalue comes from the \(X-Y\) pairing on-site as shown in Fig.\ref{fig:pair_suscep RPA}. Upon RPA, above the critical temperature \(T_C\), we find that the peak in the center of mass momentum is at the Gamma point \(\boq = \boldsymbol{\Gamma}\). However, we also note that there are secondary peaks at the \(\boldsymbol{M}\) points as well, which explains the strong competition between a \((\pi, \pi)\) PDW versus a simple \(d\)-wave pairing. In future works, one can imagine adding additional interactions which will further enhance the peak at the \(\boldsymbol{M}\) points. Lastly, since its the \(J_1\) interaction which enhances these responses, changing the sign of \(J_2\) from being ferromagnetic to anti-ferromagnetic shows minor changes in the susceptibility in the cases when \(|J_2| \ll |J_1|\).

\section{Momentum space phase diagrams}\label{appendix:kSpace}
\begin{figure}[h!]
\centering
\includegraphics[width=0.7\textwidth]{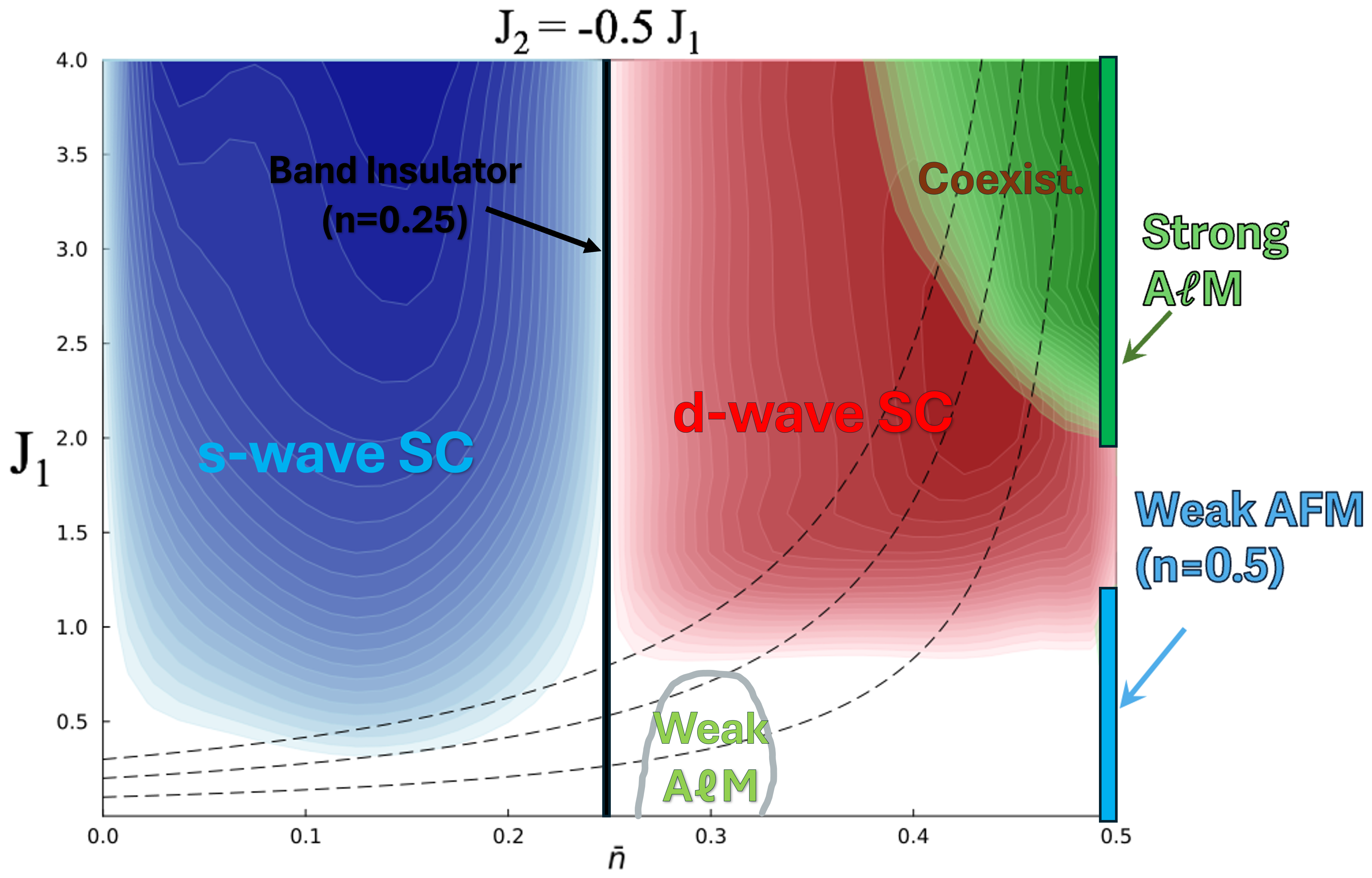}
\caption{Momentum space phase diagram for case $J_2=-0.5J_1$, assuming 2x2 diamond enlarged unit-cell. The dashed lines represent $(J_1/t)_{eff} = \frac{J_1}{t} \frac{1}{(1-2n)(1-n)}$ for values of $J_1=\{0.1,0.2,0.3\}$, where this effective exchange to hopping ratio arises from a renormalized mean field theory of the $tJ$ model as discussed in the main text.}
\label{fig:KSpace_Diag}
\end{figure}

\section{Symmetry analysis of pairing phases}\label{appendix:symmetries}
In this section, we will try to perform a symmetry analysis of various possible pairing phases seen in our phase diagram. To begin with, since we only encounter pairing within the squares of our square-octagon lattices, we will focus on such pairing states only. These \emph{intra-diamond} pairing states can be fully classified by their quantum-numbers under real-space rotation, as well as spin-space rotation or time-reversal. The order parameters for these states can be represented succinctly in real-space as (using the definition of pairing fields given above \eqref{pairing vertex})
\begin{equation}
    \label{pairing states possible}
    \Delta_{l}^{a}(\t) = \sum_{\br}\left(\expval{\D_{\br}^{a, 12}(\t)} + e^{i l \frac{\pi}{2}}\expval{\D_{\br}^{a, 23}(\t)} +  e^{i l 2 \frac{\pi}{2}}\expval{\D_{\br}^{a, 34}(\t)} +  e^{i l 3\frac{\pi}{2}}\expval{\D_{\br}^{a, 41}(\t)}\right)\,,
\end{equation}
where \(l \in \{0, \pm 1, 2\}\) represents the \emph{angular-momentum} of the pairing state (represented by \(s\), \(p_{\pm}\), and \(d\) respectively). Furthermore the spin of the paired state is denoted by \(a\), where \(a = y\) represents singlets, while \(a = x, z, 0\) represents triplets.
Now, the symmetry operations present in our hamiltonian \eqref{tJHam}, that we want to classify our pairing states by are rotation \(\mathcal{C}_4\); time-reversal \(\mathcal{T}\); inversion \(\mathcal{I}\); mirror along \(x\), \(\mathcal{M}_x\) ; mirror along \(y\), \(\mathcal{M}_y\); and spin-rotation symmetry \(\mathcal{R}_{\mathcal{S}}\).
Let us start with the simplest of these: spin-rotation symmetry. We have already categorized the pairing states according to their total spin into singlets \(S=0\) and triplets \(S=1\). We can subsequently use \(S_z^{tot}\) quantum numbers to further calssify the triplets themselves. Please note that even in the symmetry broken \alm\! or AFM phases, we have a residual \(U(1)\) symmetry corresponding to spin-rotations around the \(z\)-axis. Hence \(S_z^{tot}\) is still a good quantum number even in the magnetically ordered states to classify possible coexisting pairing phases. Consequently, the triplet states are additionally split into \(S_z^{tot} = 1\) for \(\D_l^{z}(\t) + \D_l^{0}(\t)\), \(S_z^{tot} = 0\) for \(\D_l^{x}(\t)\), and \(S_z^{tot} = -1\) for \(\D_l^{0}(\t) - \D_l^{z}(\t)\).

Since we never encounter \(p-\)wave pairing states, or finite \(S_z^{tot}\), in our phase diagram, we will only be focusing on \(s\) and \(d-\)waves singlet and triplet with \(S_z^{tot} = 0\). Let us define the action of all the aforementioned symmetry operations on the electrons, and follow that with how the various order parameters behave under such symmetries.
\begin{itemize}
    \item \textbf{Rotation} : \(c_{\br, 1, \a}\rightarrow c_{\mathcal{R}.\br, 2, \a}\,, c_{\br, 2, \a}\rightarrow c_{\mathcal{R}.\br, 3, \a}\,, c_{\br, 3, \a}\rightarrow c_{\mathcal{R}.\br, 4, \a}\,, c_{\br, 4, \a}\rightarrow c_{\mathcal{R}.\br, 1, \a}\,,\) where \(\mathcal{R}.\br\) represents the rotated unit cell position.
    \item \textbf{Time-reversal} : \(c_{\br, i, \a}\rightarrow i\s^{y}_{\a\b}\cdot c_{\br, i, \b}\) followed by a complex conjugation operation \(\mathcal{K}\).
    \item \textbf{Inversion} : \(c_{\br, 1, \a}\rightarrow c_{-\br, 3, \a}\,, c_{\br, 2, \a}\rightarrow c_{-\br, 4, \a}\,, c_{\br, 3, \a}\rightarrow c_{-\br, 1, \a}\,, c_{\br, 4, \a}\rightarrow c_{-\br, 2, \a}\,,\).
    \item \textbf{Mirror along \(x\)} : \(c_{\br, 1, \a}\rightarrow c_{\mathcal{M}_x.\br, 1, \a}\,, c_{\br, 2, \a}\rightarrow c_{\mathcal{M}_x.\br, 4, \a}\,, c_{\br, 3, \a}\rightarrow c_{\mathcal{M}_x.\br, 3, \a}\,, c_{\br, 4, \a}\rightarrow c_{\mathcal{M}_x.\br, 2, \a}\,,\) where \(\mathcal{M}_x.\br\) represents the mirrored along \(x\) unit cell position.
    \item \textbf{Mirror along \(y\)} : \(c_{\br, 1, \a}\rightarrow c_{\mathcal{M}_y.\br, 3, \a}\,, c_{\br, 2, \a}\rightarrow c_{\mathcal{M}_y.\br, 2, \a}\,, c_{\br, 3, \a}\rightarrow c_{\mathcal{M}_y.\br, 1, \a}\,, c_{\br, 4, \a}\rightarrow c_{\mathcal{M}_y.\br, 4, \a}\,,\) where \(\mathcal{M}_y.\br\) represents the mirrored along \(y\) unit cell position.
\end{itemize}
We list the corresponding quantum numbers of the pairing states that we are interested in Table.\ref{table:symmetries} under the aforementioned symmetries.

\begin{table}
\label{table:symmetries}
\begin{tabular}{ |c|c|c|c|c|c|c| } 
 \hline
 pairing state & Order Parameter & \(\mathcal{C}_4\) & \(\mathcal{T}\) & \(\mathcal{I}\) & \(\mathcal{M}_x\) & \(\mathcal{M}_y\)\\ 
\hline
s-wave singlet & \(\D_0^{y}\) & 1 & 1 & 1 & 1 & 1 \\
\hline
d-wave singlet & \(\D_2^{y}\) &-1 & 1 & 1 &-1 &-1 \\
\hline
s-wave triplet & \(\D_0^{x}\) & 1 &-1 & 1 &-1 &-1 \\
\hline
d-wave triplet & \(\D_2^{x}\) &-1 &-1 & 1 & 1 & 1 \\
\hline
\end{tabular}
\caption{Quantum numbers of possible pairing states under symmetries of the Hamiltonian.}
\end{table}

\end{document}